\documentclass[11pt]{article}
\usepackage{fullpage,parskip}

\usepackage[round]{natbib}

\usepackage[utf8]{inputenc}
\usepackage[T1]{fontenc}

\usepackage{amsmath,amssymb,amsfonts}
\usepackage{amsthm}
\usepackage{mathtools}
\usepackage{bm}
\usepackage{bbm}
\usepackage{mleftright}

\usepackage{graphicx}
\usepackage{booktabs}
\usepackage{multirow}
\usepackage{wrapfig}
\usepackage{float}
\usepackage{caption}
\usepackage{subcaption}

\usepackage{algorithm}
\usepackage{algpseudocode}

\usepackage{enumitem}
\usepackage{nicefrac}
\usepackage{microtype}
\usepackage{xcolor}
\usepackage{comment}
\usepackage{url}

\usepackage[
    hidelinks,
    colorlinks=true,
    citecolor=black!30!blue,
    linkcolor=black!10!red
]{hyperref}

\usepackage{thmtools}

\newtheoremstyle{sltheorem}
    {}          
    {}          
    {\slshape}  
    {}          
    {\bfseries} 
    {.}         
    { }         
    {}          

\theoremstyle{sltheorem}

\newtheorem{definition}{Definition}

\newtheorem{corollary}{Corollary}
\newtheorem{theorem}{Theorem}
\newtheorem{lemma}{Lemma}

\newcommand{\Real}{\mathbb{R}}
\newcommand{\norm}[1]{\left\|{#1}\right\|}
\newcommand{\Exp}[2]{E_{#1}\left[{#2}\right]}

\newcommand{\Probd}[2]{P_{#1}\left[{#2}\right]}

\usepackage{xspace}
\usepackage{bbm}
\usepackage{dsfont}
\usepackage{bm}

\newcommand{\xv}{{\bm x}}

\newcommand{\yv}{{\bm y}}

\newcommand{\cv}{{\bf c}}

\let\H\relax
\DeclareMathOperator\H{H}

\def\textiid{i.i.d.\@\xspace}
\newcommand\iid{\ifmmode\text{ i.i.d. } \else \textiid \fi}

\title{Pairwise Distortion Distribution with Graph-Based Importance Sampling for Quantization Error and Intrinsic Dimension Estimation}

\author{Xi Chen, Carl Nuzman, Shirin Jalali}

\date{}

\begin{document}

\maketitle

\renewcommand\thefootnote{}
\footnotetext{
Xi Chen and Shirin Jalali are with the Department of Electrical and Computer Engineering,
Rutgers University, Piscataway, NJ 08854 USA
(e-mail: xi.chen15@rutgers.edu; shirin.jalali@rutgers.edu).}

\footnotetext{
Carl Nuzman is with the Nokia Bell Labs, Murray Hill, NJ 07974 USA
(e-mail: carl.nuzman@nokia-bell-labs.com).
}

\footnotetext{
This work was done during Xi Chen's internship at Nokia Bell Labs.
}

\begin{abstract}
Quantifying the dimensionality of data is a fundamental need in machine learning and signal processing applications such as representation learning, data compression, and inverse problems, where the extrinsic dimension in which the data is presented can be very different from the intrinsic dimension that determines the operational performance. Various concepts, such as R\'enyi information dimension, rate-distortion dimension, and quantization dimension are useful here. Despite their theoretical significance, estimating these measures are challenging given their asymptotic definitions. Recent research has shown that analyzing the pairwise distortion distribution (PDD) of a data source can 
provide a tractable lower bound on the distortion of random codebook quantization, from which a tractable notion of intrinsic dimension can be derived.
In this paper we extend this line of research in three main ways. First, we develop a related upper bound on distortion of random codebook quantization. Second, we improve the computability of both the lower and upper bounds, using a graph-based adaptive importance sampling approach. Third, we leverage the bounds to define new notions of intrinsic dimension which form upper and lower bounds on the R\'enyi information dimension, rate-distortion dimension, and quantization dimension, and which can be empirically estimated from unknown data sources. We illustrate the concepts by analyzing several artificial and empirical distributions.
\end{abstract}

\section{Introduction}\label{sec:intro}
Developing measures of complexity that capture the intrinsic  dimension of continuous-valued sources is a  fundamental problem with applications in inverse problems, data representation and  compression.  While  for discrete sources, Shannon entropy provides a canonical and widely-adopted notion of complexity \citep{shannon1948mathematical}, no such broadly accepted notion of complexity exists for continuous-valued sources. Several notions of dimension have been developed to quantify different aspects of the complexity of continuous-valued sources, including the rate-distortion dimension (RDD)~\citep{kawabata1994rate}, R\'enyi information dimension (RID)~\citep{wu2010renyi}, Minkowski dimension and the Hausdorff dimension \citep{hausdorff1918dimension,falconer2013fractal}. These notions of complexity capture different aspects of the source complexity and have  different operational applications. For instance, the RID of a source characterizes the minimum sampling rate required  for  almost-lossless recovery from under-sampled linear measurements \citep{wu2010renyi, wu2012optimal, donoho2013information, jalali2017universal, rezagah2017compression}.

Despite their  theoretical and operational significance, estimating these  measures of complexity from data remains highly challenging. For instance, to estimate the RDD of a source, one needs to accurately estimate its corresponding rate-distortion function $R(D)$ in the small distortion  regime ($D\to 0$), which is particularly difficult given finite data. Similar challenges arise in estimating the other mentioned notions of complexity. 

An alternative notion of complexity arises from  random codebook quantization (RCQ). RCQ is a classic compression method that is based on forming a random codebook consisting of $L$ codewords independently drawn from a distribution $P_C$. To encode each input vector, an RCQ code assigns it to the closest codeword in the codebook.  An RCQ-based dimension can be defined as $d_Q(X) = \lim_{L \rightarrow \infty} - \frac{\log L}{\log \mu_L(P_X, P_C)}$~\citep{zador1982asymptotic, zhu2011asymptotic, zhu2012note, zhu2016asymptotics}, where $\mu_L(P_X, P_C)$ measures the expected  distortion achieved by an RCQ with $L$ codewords distributed as $P_C$, and the source distribution denoted as $P_X$.  Directly estimating $d_Q(X)$ is prohibitive as it requires estimating the performance of RCQ over increasingly large codebooks. However, recent work \citep{nuzman2026pairwise} connects the defined  dimension and  pairwise distortion distribution (PDD). PDD is defined as the density function of the distance between two randomly  drawn source samples.  The key advantage of this connection is that unlike RCQ-based dimension, estimation of PDD does not require generating a large codebook.

Our goal in this paper is to develop a computationally efficient and theoretically grounded method for estimating source complexity. To achieve this goal, we characterize   lower and upper bounds on RCQ distortion that can be estimated using PDD. The PDD itself can be estimated from  pairs of source samples, without knowing the source distribution.  We also define the PDD dimension, and connect the PDD dimension with the RID, RDD, and quantization dimension.

Accurate estimation of PDD in the small-distance regime remains challenging. Since the event of a randomly drawn pair satisfying this distance requirement is a rare event, a naive estimator will suffer from  high variance. To address this issue, we propose a graph-based importance sampling strategy. Given $N$ samples, the complexity of sampling on the constructed minimum spanning tree through a proposed random walk is typically $O(N \log N)$, which makes the method scalable to large datasets.

We evaluate the proposed method on different sources with known and unknown distributions. Our results establish that  PDD provides informative bounds on RCQ performance across different bitrate regimes and enables estimation of  intrinsic dimensions of sources. Moreover, the proposed method reveals the smallest and largest intrinsic dimensions of a source when it has a mixture of dimensionalities. Finally we show that the graph-based importance sampling approach considerably improves the effectiveness of PDD estimation. 

\vspace{-2mm}

Our main contributions are
\begin{itemize}
    \item We derive the lower and upper bounds on RCQ that can be computed from PDD.
    \item We define PDD dimension and establish its connections to RID, RDD, and quantization dimension.
    \item We propose a graph-based adaptive importance sampling method for effective PDD estimation in small distortion regime. The RCQ lower and upper bounds, as well as the PDD dimensions computed from the estimated PDD are evaluated on sources with known and unknown distribution.
\end{itemize}

\section{Codebook Quantization} \label{sec:RCQ}
Vector quantization (VQ) maps a sequence of continuous or discrete vectors into a digital sequence for communication and storage \citep{lloyd1982least, gray1984vector,gersho2012vector}, where the codebook is constructed for being used in VQ. The codebook quantization error is essential to understanding the rate-distortion performance of compressing the data. Let $\xv \in\mathbb R^n$ be a random vector, $\xv \sim P_X$, $\cv \in\mathbb R^n$ be a random codeword, $\cv \sim P_C$. Given a codebook consisting of $L$ codewords, we denote the codebook as 
\begin{align}
    \mathcal{C}^L=\{\cv_1,\ldots,\cv_L\}, \quad \cv_{\ell} \overset{\mathrm{i.i.d.}}{\sim}P_C, \quad \ell=1,\cdots L,
\end{align}
where $\cv_{\ell}$ is the $\ell$-th codeword. The expected quantization error of given $\xv$ induced by VQ using $\mathcal{C}^L$ is
\begin{align}\label{eq:RCQ_def}
    \mu_L(P_X,P_C)
    =
    \mathbb{E}_{X, C}
    \left[
    \min_{1\le \ell\le L}\rho(\xv,\cv_\ell)
    \right],
\end{align}
where $\rho(\xv,\cv_\ell)=\|\xv-\cv_\ell\|_p^\nu$ is the distortion measure, $\nu>0$. The minimum distortion is found in $\mathcal{C}^L$. Given realization $\xv$, the distortions $\rho(\xv,\cv_\ell)$, $\ell=1,\cdots,L$, are i.i.d. random variables. Define the conditional distortion CDF as $F_C(r;\xv) = \mathbb P_{\cv \sim P_C}\big(\rho(\xv, \cv)\le r\big)$,
then $\mathbb{P}\big(
\min_{1\le \ell\le L}\rho(\xv,\cv_\ell)>r
\big) = \big(1-F_C(r;\xv)\big)^L$. Using the tail-integral identity for nonnegative random variables, \eqref{eq:RCQ_def} can be re-written as 
\begin{align}\label{eq:RCQ_def_2}
    \mu_L(P_X,P_C) = \mathbb E_X \left[ \int_0^\infty \big(1-F_C(r;\xv)\big)^L\,dr
\right].
\end{align}
When $L$ is large, the integral is dominated by behavior of $F_C(r;\xv)$ for small radius $r$, which in turn reflects increasingly local behavior of the distributions $P_X$ and $P_C$.
An important special case of RCQ is to draw the codeword from the same distribution as the data, $P_C = P_X$. In many cases, particularly if the dimension is not too small, the codebook distribution minimizing say mean squared error distortion is close to the data distribution. Working with $P_X$ is also convenient for unknown, empirical distributions.  In this paper we will focus on the case $P_C=P_X$, although many of the methods and concepts can be extended to a more general setting.
One useful object for studying RCQ was introduced in~\citep{nuzman2026pairwise}. 
\begin{definition}[Pairwise distortion distribution~\citep{nuzman2026pairwise}] \label{def:1}
Let $\xv \sim P_X$ be a random vector on $\mathcal{X}$, and let
$\rho:\mathcal X\times\mathcal X\to[0,\infty)$ be a distortion measure. With $F_X(r;\xv) = \mathbb P_{\yv \sim P_X}\big(\rho(\yv, \xv)\le r\big)$, the pairwise distortion distribution (PDD) is defined as
\[
F(r)=\mathbb E_X[F_X(r;\xv)] =
\mathbb P_{\xv_1,\xv_2\sim P_X}\big(\rho(\xv_1,\xv_2)\le r\big).
\]
\end{definition}
This distribution can be used to express a simple lower bound on RCQ error. In Section~\ref{sec:RCQ_bounds}, we introduce a related function and use it to define an upper bound of RCQ. In Section~\ref{sec:PDD_dimension}, we define notions of intrinsic dimension based on these RCQ bounds and show how they relate to other definitions of intrinsic dimension. 
In Section~\ref{sec:PDD_estimation}, we introduce methodology for empirically estimating the PDD from data, and in Section~\ref{sec:PDD_estimation_IS}, propose an importance sampling method to improve the scalability of this computation. Numerical experiments with the methods are reported in  Section~\ref{sec:Simulation}.

\section{PDD-based RCQ Bounds} \label{sec:RCQ_bounds}
We derive the PDD-based RCQ bounds in this section. 
Detailed proofs of the results are given in Section~\ref{appendix:proofs}. We begin with a definition.
\begin{definition}[Averaged inverse conditional PDD]\label{def:2}
For $q\in[0,1]$, define the inverse of $F(r;\xv)$ as
$G(q;\xv)=\sup\{r\ge 0:F(r;\xv)\le q\}$, with $\sup\emptyset=0$. 
The averaged inverse conditional PDD is $G(q)=\mathbb E_X[G(q;\xv)]$.
\end{definition}
Taking $P_C=P_X$, we modify the notation of RCQ as $\mu_Q(L)$ to emphasize the dependence on $L$, and to differentiate it from related function $\mu_O(L)$ and $\mu_U(L)$ defined later. We derive the lower and upper bound of the RCQ distortion in terms of $F(r)$ and $G(q)$ in the main theorem. The lower bound result was already given in \citep{nuzman2026pairwise}. 
\begin{theorem}[PDD-based RCQ bounds] \label{theorem:1}
Let $P_C=P_X$, $F(r)$ and $G(q)$ as defined in Definition~\ref{def:1},\ref{def:2}, with $\int_0^\infty(1-F(r))\,dr < \infty$ and $G(q)<\infty$ for $q\in[0,1]$. The RCQ is bounded as
\[
\mu_Q(L) \ge \underline{\mu}_Q(L) = \int_0^\infty (1-F(r))^L\,dr,
\]
\[
\mu_Q(L) \!
\le
\overline{\mu}_Q(L) \! = \!\!\!\!
\inf_{t\in[0,1]}
\left[
t^{L-1}\mu_Q(1) \!+\!(1-t^L)G(1-t)
\right].
\]
\end{theorem}
The bounds coincide for a single code point: $\underline{\mu}_Q(1)=\mu_Q(1)=\overline{\mu}_Q(1)$.
The lower and upper bounds in Theorem~\ref{theorem:1} are particularly useful for understanding the rate-distortion performance at different bitrate regimes of an unknown source, when the exact RCQ is hard to obtain. For a general source distribution, $F(r)$ and $G(q)$ can be estimated by sampling the pairs from the source distribution, as explained in Section~\ref{sec:PDD_estimation}. Any value of $t\in[0,1]$ gives a valid upper bound; the choice of $t$ can be optimized numerically given $G(q)$.

The behavior of $F(r)$ and $G(q)$ for small $r$ and $q$ gives insight into the asymptotic behavior of $\mu_Q(L)$ for large $L$, as indicated by these two lemmas.

\begin{lemma}\label{lemma:pdd_lb}
Let $F(r)$ satisfy $\lim_{r\to 0} \frac{\log(F(r))}{\log(r)} = d \geq 0$. Then
$\underline{\mu}_Q(L) $ satisfies $\lim_{L\to\infty} \frac{\log(1/L)}{\log(\underline{\mu}_Q(L))} = d$.
\end{lemma}
\begin{lemma}\label{lemma:pdd_ub}
Let $G(q)$ satisfy $\lim_{q\to 0} \frac{\log(q)}{\log(G(q))} = d \geq 0$. Then
$\overline{\mu}_Q(L)$ satisfies $\lim_{L\to\infty} \frac{\log(1/L)}{\log(\overline{\mu}_Q(L))} = d$.
\end{lemma}
Likewise, explicit bounds on the functions $F(r)$ and $G(q)$ yield convenient explicit bounds on RCQ error.
\begin{corollary}[\citep{nuzman2026pairwise}]\label{cor:F_bound}
Let $\xv \sim P_X$ on $\mathbb{R}^{n}$ and suppose $F(r) \leq \beta r^{\alpha}$ for $\alpha>0$, $\beta > 0$. Then
$$
\mu_Q(L) \geq \underline{\mu}_Q(L) \geq \alpha^{-1} \beta^{-1/\alpha} B(L+1,\alpha^{-1}),
$$
where B is the Beta function.
\end{corollary}

\begin{corollary}\label{cor:G_bound}
Let $\xv \sim P_X$ on $\mathbb{R}^{n}$ and suppose that $G(q) \leq \gamma q^{1/\alpha}$ for some $\alpha>0$, $\gamma > 0$. Then
\begin{align*}
\overline{\mu}_Q(L)
\leq
\mu_Q(1)
\left(1-s_L^{-1}\right)^{L-1}  +  \gamma s_L^{-1/\alpha},
\end{align*}
where $s_L=(\alpha L)/\log(L)$.
\end{corollary}
The behavior of the functions $F(r)$ and $G(q)$ near $0$ have an explicit description when $\xv$ has a sufficiently smooth density on $\Real^n$.
Let $V_{p,n}(1)$ denote the volume of the ball of radius $1$ under the $p$-norm on $\Real^n$, in which case the ball of radius $r$ has volume $V_{p,n}(1)r^n$.
\begin{theorem}\label{theorem:cts-case}
    Suppose that $\xv$ has absolutely continuous distribution on $\Real^n$ with density $p(\xv)$, and that $p:\Real^n\to\Real$ is a uniformly continuous function. Let the distortion function be $\rho(\xv,\yv)=\|\xv-\yv\|_p^\nu$ for $\nu>0$.
    Then
    $$
    \lim_{r\to 0} F(r)r^{-n/\nu} = E_X\left[ p(\xv) \right] V_{p,n}(1)  ,
    $$
    and 
    $$
    \lim_{q\to 0} G(q)q^{-\nu/n} = E_X\left[ p(\xv)^{-\nu/n} \right] V_{p,n}(1)^{-\nu/n}, 
    $$
    if the respective expectations are finite.
\end{theorem}
The proof is based on the approximation $F(r;x) \approx p(\xv) V_{p,n}(1) r^{n/\nu}$ for small $r$.

 In the case of a probability density on $\Real^n$, and distortion with $\nu=1$, it is also known that $\mu_Q(L) \sim L^{-1/n}$~\citep{zador1982asymptotic}.
Thus, given a distribution on $\Real^n$ such that $F(r) \sim r^d$, $G(q)\sim q^{1/d}$, and $\mu_Q(L) \sim L^{-1/d}$, we can say that it "behaves like"  a continuous distribution on $\Real^d$, and we can define $d$ as its intrinsic dimension. We develop these ideas in the next section.

\section{PDD dimension} \label{sec:PDD_dimension}
In this section, we define new intrinsic dimensions and relate them to the RDD, RID, and quantization dimension discussed in Section~\ref{sec:intro}. 

For an arbitrary distortion function, we have denoted the expected distortion using RCQ with $L$ codewords by $\mu_Q(L)$. We likewise denote by $\mu_O(L)$ the distortion of the optimal codebook, and by $\mu_U(L)$ the distortion achievable by source coding over unlimited block lengths, where for distortion-rate function $D(R)$, we have $\mu_U(L) = D(\log L)$. We have $\underline{\mu}_Q(L) \leq \mu_Q(L) \leq \overline{\mu}_Q(L)$, for our RCQ bounds, and we have $\mu_U(L) \leq \mu_O(L) \leq \mu_Q(L)$ by definition.
For an arbitrary distortion measure, we denote the \emph{uncalibrated} RCQ dimension as \begin{equation}\label{eq:def_rcq_dim_uncal}
\tilde{d}_Q(X) = \lim_{L\to\infty} \frac{\log(L)}{\log\left( 1/\mu_Q(L)\right)}.
\end{equation}
For distortion $\rho(\xv,\yv)=\| \xv-\yv\|_p^\nu$, $\nu>0$, we define the \emph{calibrated} dimension as $d_Q(X) = \nu \tilde{d}_Q(X)$. Computing dimension via the calibrated definition yields the same value, regardless of the exponent $\nu$ defining the distortion, while the uncalibrated definition has the benefit of working for more general distortion functions. 
Here we will stick with distortion measures of the form  $\| \xv-\yv\|_p^\nu$ and calibrated definitions of dimension - but many of the results and methods extend naturally to general distortion measures.
Similarly to (\ref{eq:def_rcq_dim_uncal}), we define $d_O(X),\tilde{d}_O(X),d_U(X),\tilde{d}_U(X)$ based on $\mu_O(X)$ and $\mu_U(X)$.


The quantization dimension $d_Q(X)$ coincides with the definition in~\citep{zador1982asymptotic}, and 
$d_U(X)$  is the same as the RDD defined in \citep{kawabata1994rate}, since by change of variables
\begin{equation}\label{eq:rdd_formula}
d_U(L) = \! \nu \lim_{L\to\infty} \frac{\log(L)}{\log(1/\mu_U(L))} = \nu \lim_{D\to 0} \frac{R(D)}{\log(1/D)}.
\end{equation}
The dimensions $d_I, d_U, d_O$, and $d_Q$ are typically hard to characterize, especially from empirical observations of unknown distributions. However, we can define notions of dimension, based on $F(r)$ and $G(q)$, which bound these notions of dimension, and which are easier to estimate empirically.

\begin{definition}[PDD-based dimension] \label{def:3}
Given distortion of the form  $\rho(\xv,\yv)=\| \xv-\yv\|_p^\nu$, the calibrated lower and upper PDD dimensions are
\begin{align*}
    \underline{d}_Q(X)= \nu \lim_{r \to 0}\frac{\log F(r)}{\log r}, \;  \overline{d}_Q(X) = \nu \lim_{q \to 0}\frac{\log q}{\log G(q)},
\end{align*}
when these limits exist, where $F(r)$ and $G(q)$ are defined in Definition~\ref{def:1}, \ref{def:2}.
\end{definition}
We note that the lower dimension $\underline{d}_Q(X)$ generalizes the correlation 
dimension~\citep{grassberger1983measuring}, $d_c(X) = \lim_{r \to 0} \frac{\log C(r)}{\log r}$, where  $C(r) = \mathbb{P}_{\xv_1,\xv_2 \sim P_X} \big( \|\xv_1 - \xv_2\| \le r \big)$.

Lemmas~\ref{lemma:pdd_lb} and \ref{lemma:pdd_ub} show that the PDD dimensions relate directly to the asymptotic behavior of the RCQ upper and lower bounds defined in Theorem~\ref{theorem:1}.
They also bound all of the other notions of dimension we have been discussing.


\begin{theorem}\label{theorem:calibrated}
For distortion measure of the form $\rho(\xv,\yv)=\| \xv-\yv\|_p^\nu$, with $1\leq p \leq \infty$ and $\nu > 0$, the calibrated dimensions satisfy
$$
\underline{d}_Q(X) \! \leq \! d_I(X) \! = \! d_U(X) \leq d_O(X) \leq d_Q(X) \leq \overline{d}_Q(X),
$$
whenever these limits exist.
\end{theorem}
See the Section~\ref{appendix:proofs} for the full proof.
The equality of RID and RDD, $d_I(X) = d_U(X)$, for the given distortion function is provided in \citep{wu2010renyi}, for example, while the bound $\underline{d}_Q(X) \leq d_I(X)$ is obtained by correcting an error in the argument originally given in \citep{nuzman2026pairwise}.


\begin{figure*}[t]
    \centering

    \begin{subfigure}{0.25\textwidth}
        \centering
        \includegraphics[width=\linewidth]{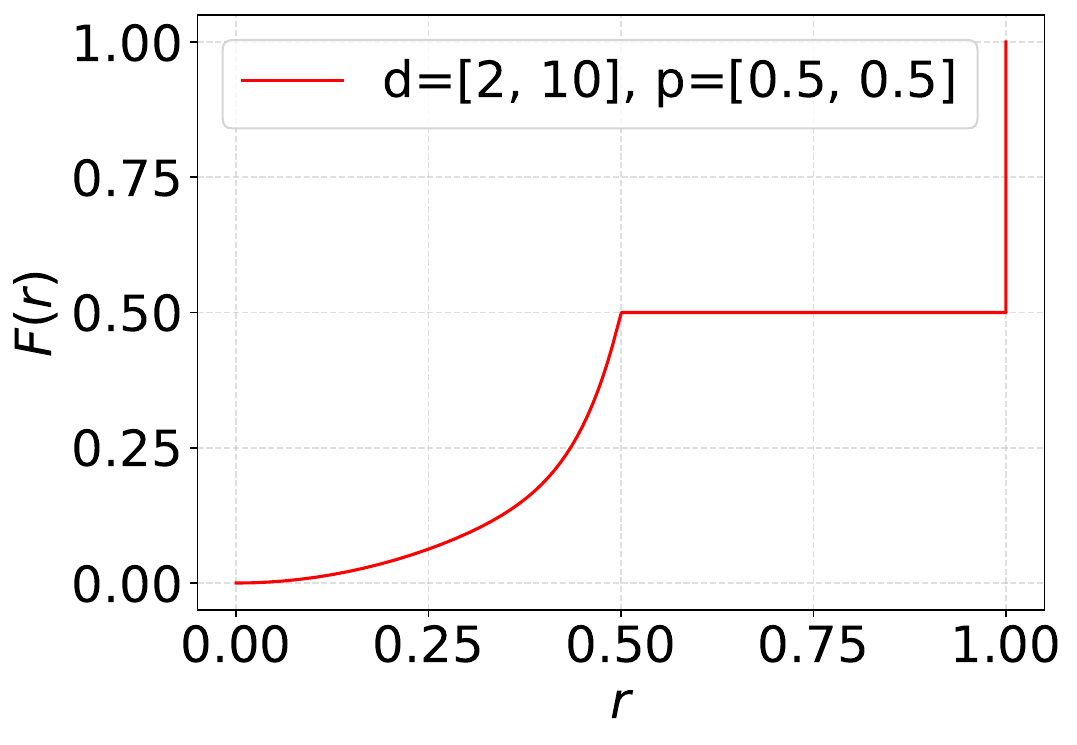}
    \end{subfigure}\hfill
    \begin{subfigure}{0.25\textwidth}
        \centering
        \includegraphics[width=\linewidth]{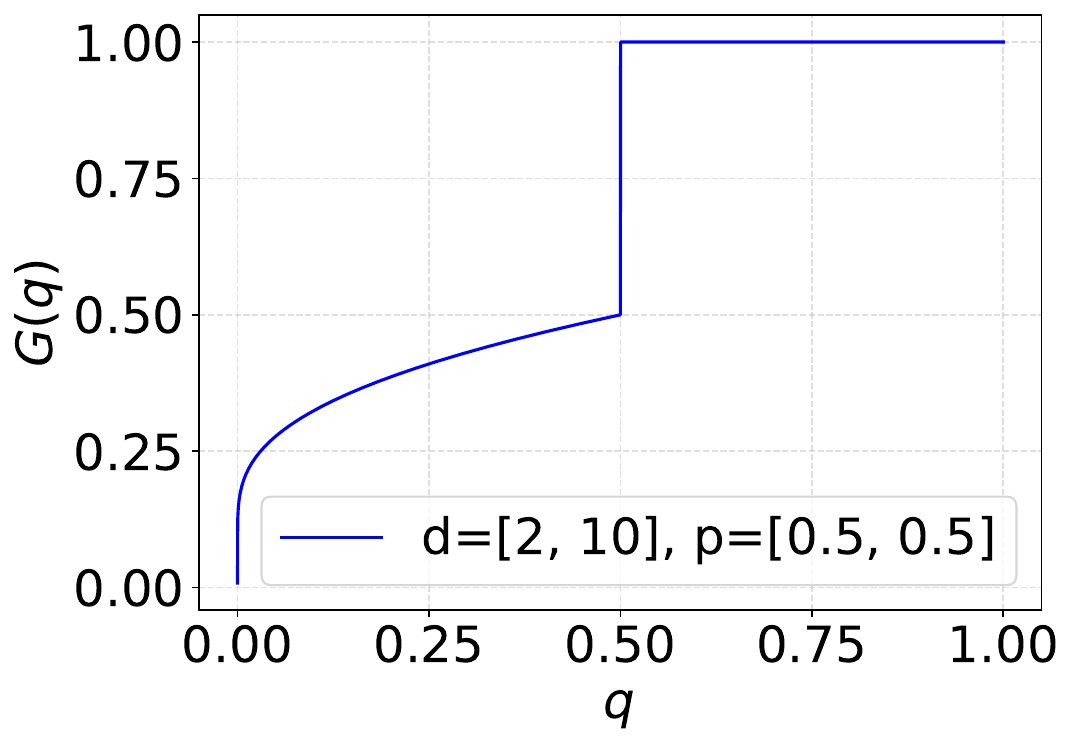}
    \end{subfigure}\hfill
    \begin{subfigure}{0.25\textwidth}
        \centering
        \includegraphics[width=\linewidth]{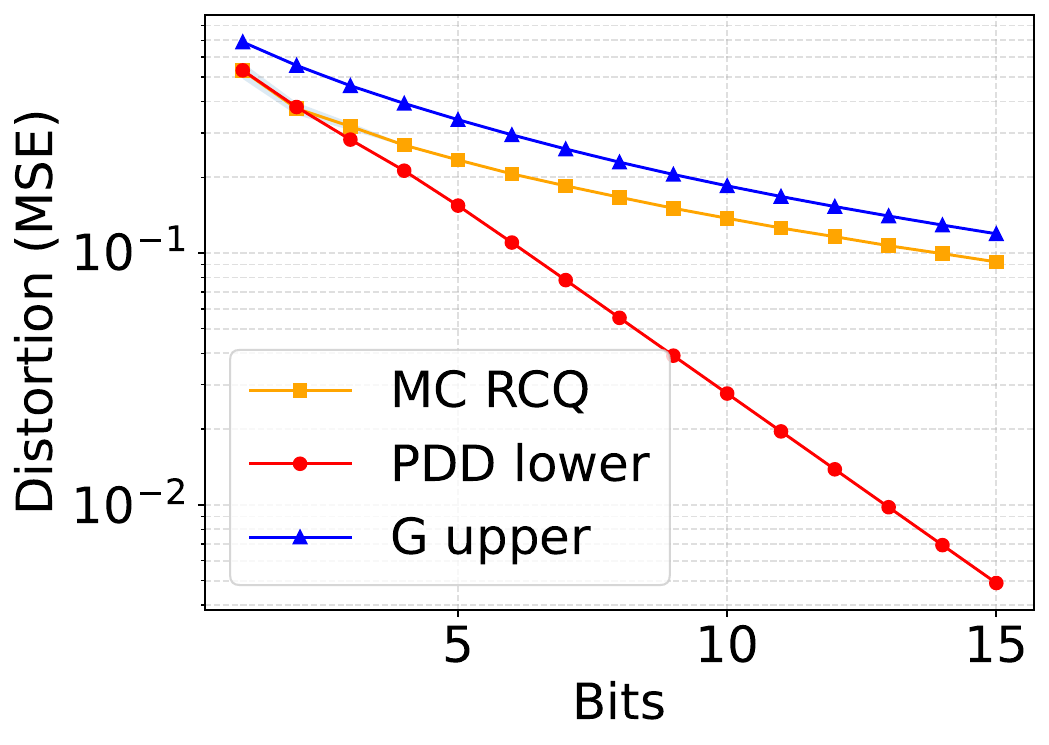}
    \end{subfigure}\hfill
    \begin{subfigure}{0.25\textwidth}
        \centering
        \includegraphics[width=\linewidth]{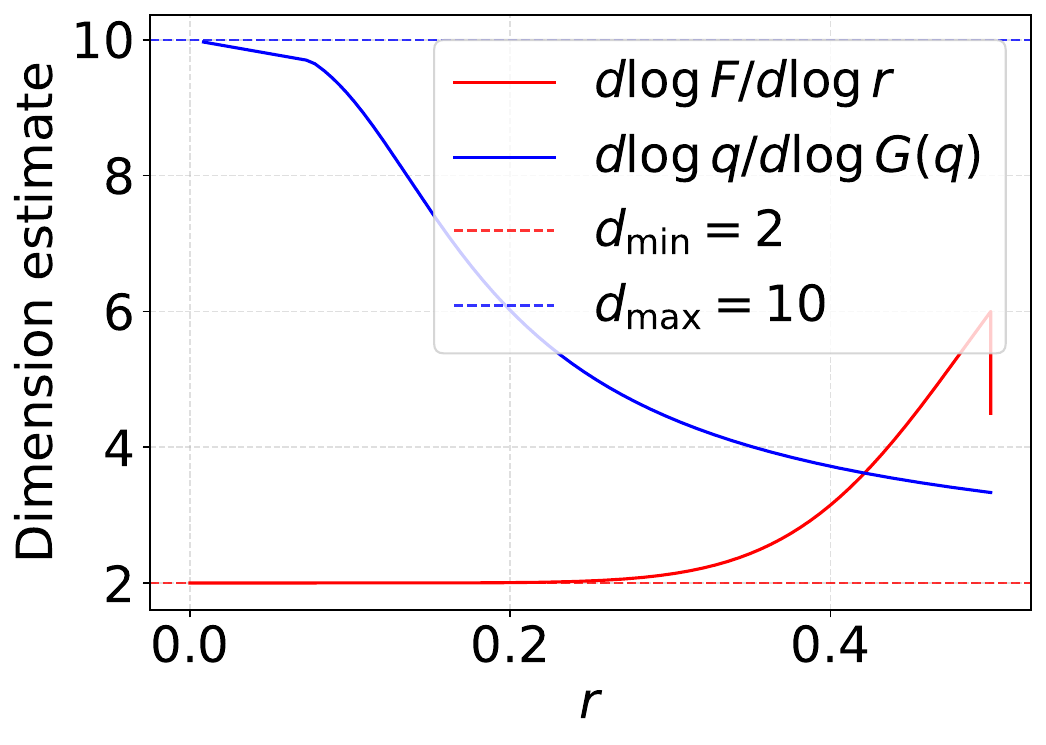}
    \end{subfigure}


    \begin{subfigure}{0.25\textwidth}
        \centering
        \includegraphics[width=\linewidth]{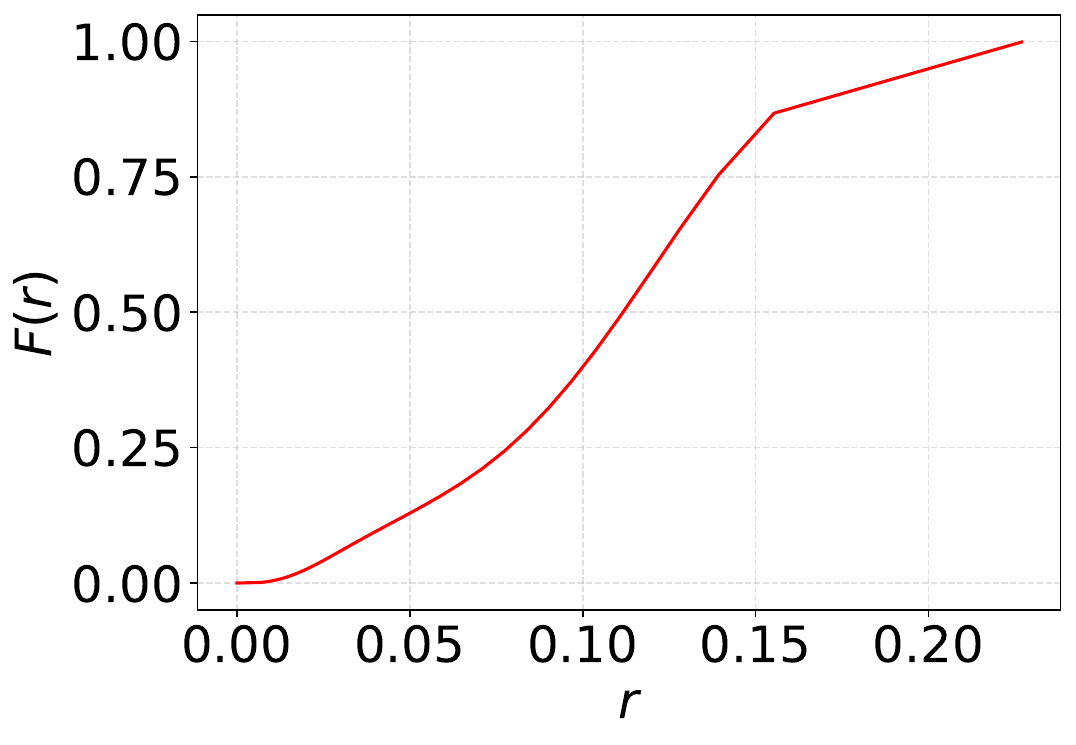}
    \end{subfigure}\hfill
    \begin{subfigure}{0.25\textwidth}
        \centering
        \includegraphics[width=\linewidth]{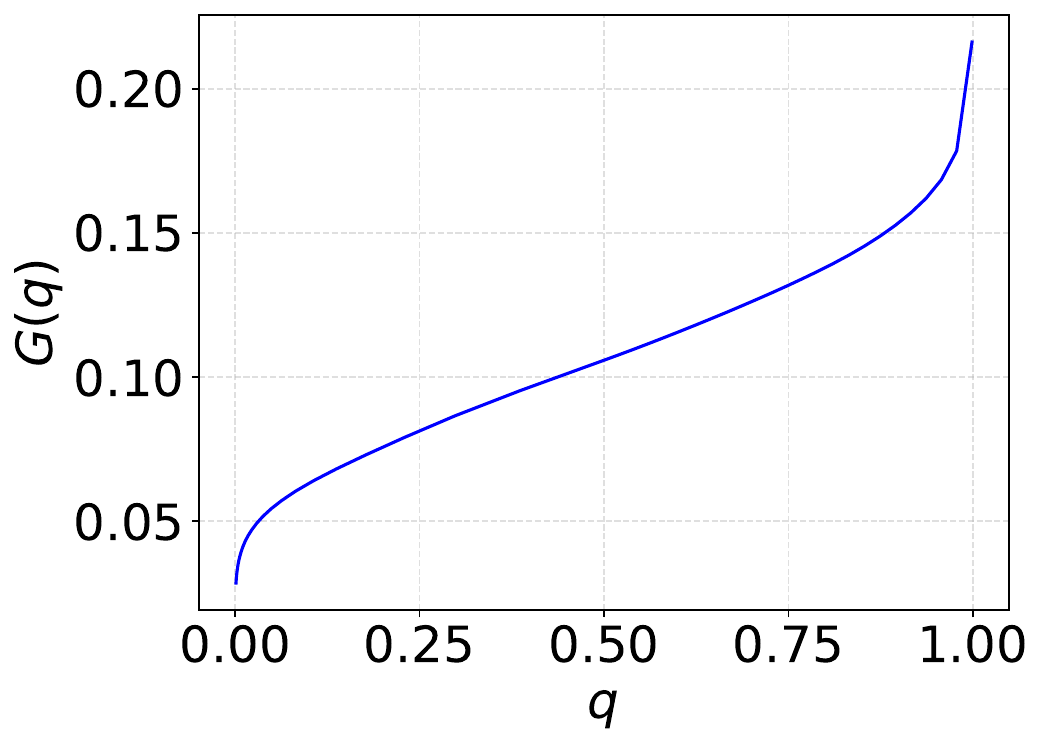}
    \end{subfigure}\hfill
    \begin{subfigure}{0.25\textwidth}
        \centering
        \includegraphics[width=\linewidth]{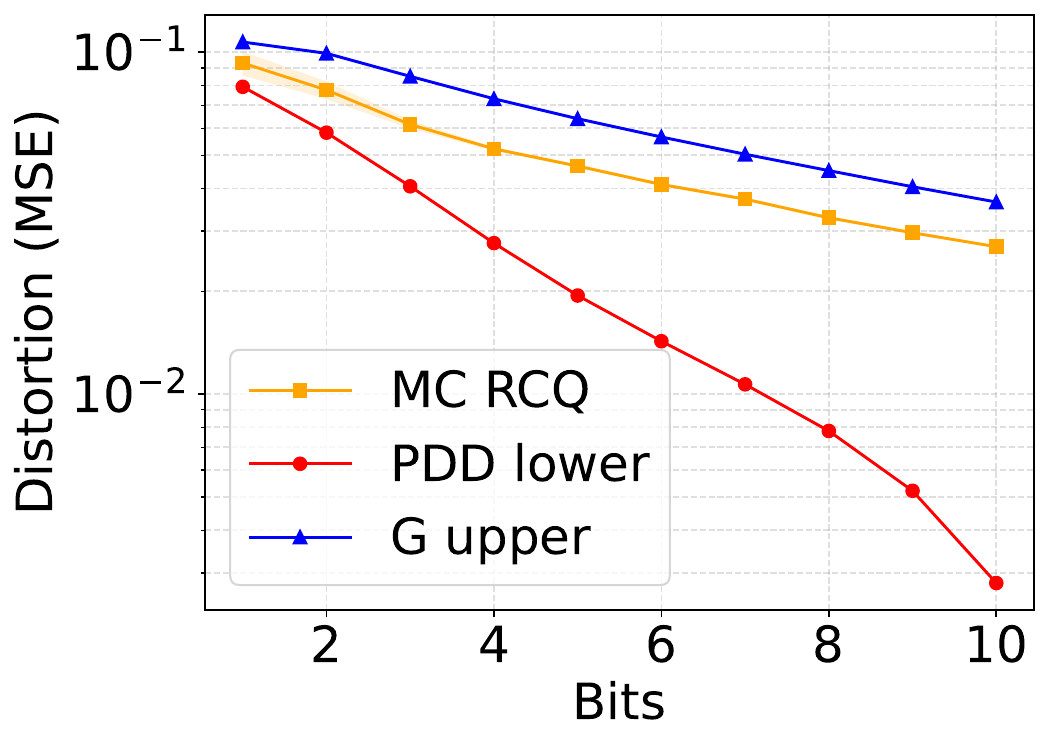}
    \end{subfigure}\hfill
    \begin{subfigure}{0.25\textwidth}
        \centering
        \includegraphics[width=\linewidth]{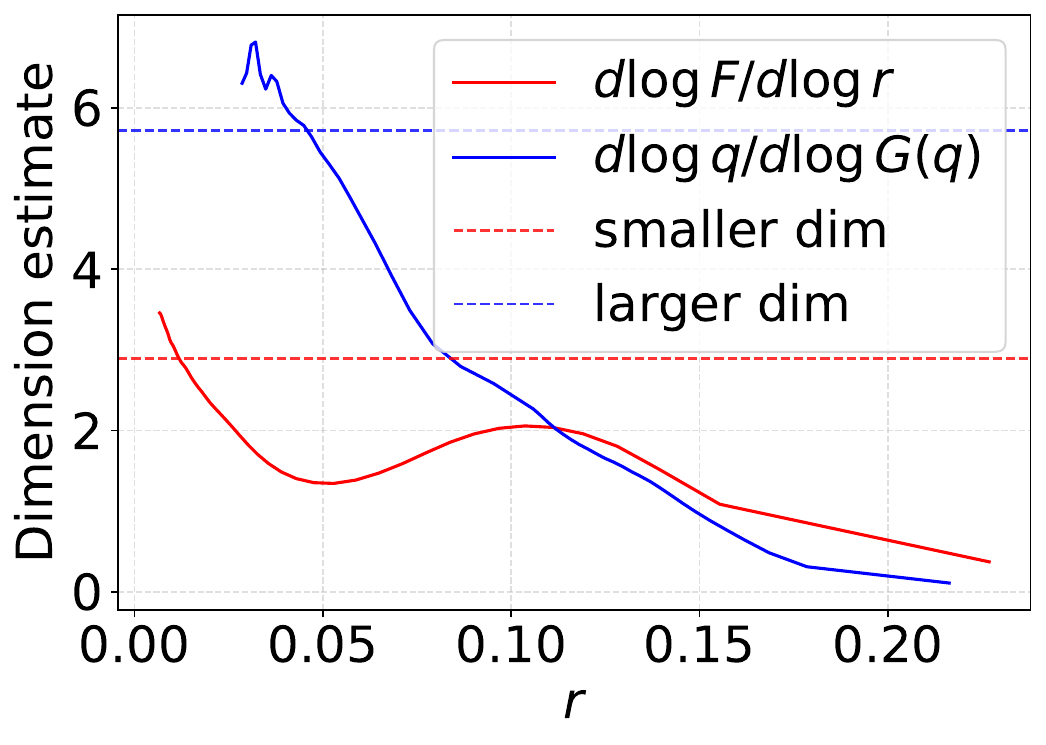}
    \end{subfigure}

    \vspace{-2mm}
    \caption{PDD-based analysis on two mixture distributions. 
    Top row: mixture of hypercubes. Bottom row: mixture of MNIST digits. 
    We show $F(r)$, $G(q)$, RCQ bounds, 
    and the estimated PDD dimension.}
    \label{fig:RCQ_hypercube_mnistmixture}
    \vspace{-4mm}
\end{figure*}

We illustrate these notions of dimension with examples where $\underline{d}_Q(X) = \overline{d}_Q(X)$ and others where $\underline{d}_Q(X) < \overline{d}_Q(X)$. 
For an absolutely continuous source on $\Real^n$, if the expectations in Theorem~\ref{theorem:cts-case} exist, then we have $\underline{d}_Q(X)=\overline{d}_Q(X)=n$, and so by Theorem~\ref{theorem:calibrated}, all of the notions of dimension coincide. Likewise, for a source on a finite set, we have $F(0)>0$ and $G(q) = 0$ for $q<F(0)$, and so $\underline{d}_Q(X)=\overline{d}_Q(X) = 0$.

A less trivial but still tractable case is the Bernouilli uniform source $X=BU$ where 
$B\sim \mathrm{Bernoulli}(p)$, $Z\sim \mathrm{Unif}[0,1]$, and
$B$ is independent of $Z$.
Here one can check that $F(0)=p^2$, yielding $\underline{d}_Q(X)=0$, while $G(q) \approx \frac{1-p}{2p} q$ for small $q$, yielding $\overline{d}_Q(X)=1$. It is well known that the RID in this case is $d_I(X)=p$, and we can also calculate $d_O(X)=d_Q(X)=1$.

Generalizing this result, when we have a mixture of distributions which have densities on manifolds of different dimensions, $\underline{d}_Q(X)$ picks the smallest of the dimensions while $\overline{d}_Q(X)$ picks the largest.
This is illustrated by the following artificial example in which $F(r)$ and $G(q)$ have simple closed form expressions.

Consider a finite mixture of unit hypercubes $X\sim \sum_{i=1}^m p_i \mathcal D_i$, where $\mathcal D_i$ is the uniform distribution on a $d_i$-dimensional
unit hypercube $\mathcal H_i$, and $\sum_i p_i=1$. 
If $\xv,\yv\in\H_i$, 
we use the torus $\ell_\infty$ distortion
\[
\rho_i(\xv,\yv)
=
\max_{1\le k\le d_i}
\min\{|x_k-y_k|,1-|x_k-y_k|\}.
\]
If $\xv \in H_i, \yv \in H_j$, $i\neq j$, we define $\rho(\xv,\yv)=1$. 
Then, for $r\le 1/2$, if $\xv \in \mathcal H_i$, $F(r;\xv)=p_i(2r)^{d_i}$. Consequently, $F(r) \sim r^{\min_i d_i}$, $G(q) \sim q^{\frac{1}{\max_i d_i}}$. Therefore, $\underline d_P(X) = \min_i d_i$, $\overline d_P(X) = \max_i d_i$. Using \eqref{eq:rid_formula}, the RID of this source  
is $d_{\mathrm I}(X) = \sum_{i=1}^m p_i d_i$. 
Thus, the PDD bounds capture the range of intrinsic dimensions present, while the RID captures "average" dimension.

The first row of Figure~\ref{fig:RCQ_hypercube_mnistmixture} illustrates the functions $F(r)$, $G(q)$ and quantization bounds for a mixture of hypercubes with $d_1=2$, $d_2=10$, and $p_1=0.5$, $p_2=0.5$.
The larger and smaller dimensions are captured by $\underline{d}_Q(X)$ and $\overline{d}_Q(X)$, respectively. The RCQ curve $\mu_Q(X)$ is sandwiched between the PDD bounds.

The second row of Figure~\ref{fig:RCQ_hypercube_mnistmixture} shows a similar analysis carried out empirically from a subset of the MNIST dataset, filtered to include only the images with labels $''1''$ and $''2''$. When analyzed separately, the data in these two classes have intrinsic dimension around $d=3$ and $d=6$, respectively. When the mixture distribution is analyzed, we get approximately $\underline{d}_Q(X)=3$ and $\overline{d}_Q(X)=6$, with RCQ behavior bounded by $\underline{\mu}_Q(X)$ and $\overline{\mu}_Q(X)$ as expected.


\section{Empirical PDD Estimation} \label{sec:PDD_estimation}
To compute RCQ bounds and bounds on intrinsic dimension for an unknown data source, we need to estimate the PDD empirically.
For this, it is helpful to  express the PDD as $F(r) = \mathbb{E}_{\xv_1,\xv_2 \sim P_X}\left[ \mathbbm{1} \left\{ \rho(\xv_1, \xv_2) \le r \right\} \right]$. 
Given a dataset of $N$ samples, we
we estimate the expectation by sampling $K$ independent pairs $(\xv_{2k-1}, \xv_{2k})$ uniformly from the dataset. The Monte Carlo (MC) based estimator is $\widehat{F}(r) = \frac{1}{K} \sum^K_{k=1} Z_k$, $Z_k = \mathbbm{1}\{\rho(\xv_{2k-1},\xv_{2k}) \le r\}$, where $\xv_{2k-1}, \xv_{2k} \overset{\mathrm{i.i.d.}}{\sim} P_X$. 


\subsection{Accuracy of MC estimation}
Assume we have $2K$ iid samples, thus $K$ pairs. Since $Z_k \sim \text{Bernoulli}(p)$, where $p=F(r)$, the estimator is unbiased $\mathbb{E}[\widehat{F}(r)] = p$, the variance of it is $\mathrm{Var}[\widehat{F}(r)] = p(1-p)/K$. So the coefficient of variation (CV) is
\[
\mathrm{CV}(\widehat{F}(r)) = \frac{\sqrt{\mathrm{Var}[\widehat{F}(r)]}}{\mathbb{E}[\widehat{F}(r)]} = \sqrt{\frac{1-p}{pK}},
\]
and its empirical value estimated from the $K$ samples is $\sqrt{(1-\widehat{F}(r)) /(\widehat{F}(r)K)}$.
In order to get a good estimate of $\widehat{F}(r)$, we need to keep $\mathrm{CV}$ small, for instance, $\mathrm{CV} \leq 1/10$ requires $pK\geq 100$.
Note that $pK$ is the expected number of non-zero values of the indicator function. So, if the number of non-zero terms is sufficiently high, we can get good estimate $\widehat{F}(r)$ using the MC, if $pK$ is small, the error bar is large. Thus, given small $r$, $F(r)$ is small, we need large $K$. Since $\widehat{F}(r)$ is an increasing function of $r$, this means, in a given experiment, there is a minimum value of $r$ below which $F(r)$ cannot be confidently estimated.

\subsection{Estimation with importance sampling}
We can adopt importance sampling (IS)~\citep{kahn1953methods} to alleviate this issue. The idea is that we twist the distribution from which the pairs are sampled, this indirectly induces a twisted distribution on the distance of pairs $\mathbb{P}(\rho(\xv_1,\xv_2) \le r)$. Denote $Q_X$ as the twisted distribution that $(\xv_1,\xv_2)$ is sampled from, each sample is corrected by $P_X(\xv_1) P_X(\xv_2)/Q_X(\xv_1,\xv_2)$ to avoid bias. So, the estimator is
\begin{align} \label{eq:pdd_estimation_is}
\widehat F_{\mathrm{IS}}(r)
=
\frac{1}{K}
\sum_{k=1}^{K}
\mathbbm{1}\{\rho(\xv_{2k-1},\xv_{2k}) \le r\} \cdot Y,
\end{align}
where $Y=P_X(\xv_{2k-1}) P_X(\xv_{2k})/Q_X(\xv_{2k-1},\xv_{2k})$, and $(\xv_{2k-1},\xv_{2k}) \sim Q_X$. When a finite dataset $\mathcal{X}$ has size of $|\mathcal{X}|=N$, then $P_X(\xv) = \frac{1}{N}$. We will introduce how to design $Q_X$ later. Before that, we analyze the MC estimator with importance sampling.

\subsection{Accuracy of importance sampling}\label{subsec:accuracy_of_IS}
Consider an individual summand of \eqref{eq:pdd_estimation_is}, which can be modeled as the product $Z=BY$ of a Bernoulli r.v. $B$, i.e., the indicator function, with a non-negative r.v. $Y$, i.e., the ratio of P to Q. We assume $Y>0$ almost surely, so that $Z=0$ only when $B=0$. This is the case as long as the twisted distribution has the same support as the true distribution. Suppose that $\mathbb{P}(B=1)=p$, then we have $\mathbb{E}[Z]=p \mathbb{E}[\tilde{Y}]$, $\mathbb{E}[Z^2]=p \mathbb{E}[\tilde{Y}^2]$, and $\mathrm{Var}(Z) = p (\mathrm{Var}(\tilde{Y}) + (1-p) \mathbb{E}[\tilde{Y}]^2)$, where $\tilde{Y}$ is the random variable obtained by conditioning $Y$ on the event $B=1$, $\tilde{Y} = Y | B=1$, i.e., the random variable $P(\xv_1)P(\xv_2)/Q(\xv_1,\xv_2)$ under the twisted distribution, conditioned on the event that $\rho(\xv_1,\xv_2)\leq r$. If each of the $K$ pairs is independent of the other pairs, and $\widehat F_{\mathrm{IS}}(r) = \frac{1}{K} \sum^K_{k=1} Z_k$, then
\begin{align} \label{eq_cv_IS}
    &\mathrm{CV}(Z)^2  = \frac{\mathrm{Var}(Z)}{\mathbb{E}[Z]^2} = \frac{1}{p} \mathrm{CV}(\tilde{Y})^2 + \frac{1-p}{p}, \nonumber\\
    &\mathrm{CV}\big(\widehat F_{\mathrm{IS}}(r)\big)^2 = \frac{1}{pK} \mathrm{CV}(\tilde{Y})^2 + \frac{1-p}{pK}.
\end{align}
The first term depends on $\mathrm{CV}(\tilde{Y})$, and the second term is the variance of the averaged $K$ independent $B's$.
Without importance sampling, we have $\mathrm{CV}(\tilde{Y})=0$ and $p \approx \widehat{F}(r)$. Thus the CV of the estimate is approximately $1/(F(r) K)$ and the number of samples $K$ required increases in inverse proportion with the probability $F(r)$ we are targeting.
With importance sampling, $p$ is no longer tied to $F(r)$, as we can make the twisted probability of the event $\rho(x_{2k-1},x_{2k})\leq r$ to be much higher than the untwisted probability of interest. 
Thus the product $pK$ can be controlled with a relatively small number of samples $K$.
However, there is an additional term $\mathrm{CV}(\tilde{Y})$. 

In the next section, we introduce a family of twisted distributions, with a single parameter that controls the degree to which nearby pairs of points are emphasized in the twisted distribution. Increasing the degree of twist generally brings up the event probability $p$ but may also eventually increase the term $\mathrm{CV}(\tilde{Y})$. Thus the optimal degree of twist involves a tradeoff between these two terms in \eqref{eq_cv_IS}.
To help analyze the tradeoff, it can be useful to empirically estimate the two terms.  The event probability can be estimated as $\hat{p} = \frac{1}{K} \sum^K_{k=1} \mathbbm{1}\{ \rho(\xv_{2k-1},\xv_{2k}) \leq r \}$. For $\tilde{Y}$, we have $\widehat{\mathbb{E}}[\tilde{Y}] = \frac{1}{K'} \sum^{K'}_{k=1} \tilde{Y}_k$, $\widehat{\mathrm{Var}}(\tilde{Y}) = \widehat{\mathbb{E}}[\tilde{Y}^2] - (\widehat{\mathbb{E}}[\tilde{Y}])^2$. Putting the pieces together,
\begin{align} \label{eq:GIS_cv}
    \widehat{\mathrm{CV}(\widehat F_{\mathrm{IS}}(r))^2} = \frac{1}{\hat{p}K} \frac{\widehat{\mathrm{Var}}(\tilde{Y})}{\widehat{\mathbb{E}}[\tilde{Y}]^2} + \frac{1-\hat{p}}{\hat{p}K}.
\end{align}

\begin{algorithm}[t]
\caption{Graph-based importance sampling}
\label{alg:graph_sampler}
\begin{algorithmic}[1]
\State Construct a $k$-nearest-neighbor graph $G=(V,E)$ on $\{\xv_i\}_{i=1}^N$ based on the distortion measure $\rho$.
\State Get a minimum spanning tree $T=(V,E_T)$ of $G$.
\For{each edge $(u,v)\in E_T$}
    \State Split $T$ at $(u,v)$ into $T_u,T_v$, $u\in T_u$, $v\in T_v$.
    \State Set $w_{u\to v}=|T_v|$ and $w_{v\to u}=|T_u|$.
\EndFor
\State Sample $i_1$ uniformly from $\{1,\ldots,N\}$.
\State Set $u=i_1$, $\mathcal V_{\mathrm{prev}}=\emptyset$.
\For{$t=1,2,\ldots$}
    \State Create $\mathcal N_t(u)=\{v:(u,v)\in E_T,\ v\notin\mathcal V_{\mathrm{prev}}\}$.
    \State Stop with probability: $\frac{1}{1+\sum_{v\in\mathcal N_t(u)}w_{u\to v}^{\tau}}$
    \State Otherwise sample $v\in\mathcal N_t(u)$ with probability
    \begin{align} \label{eq:transition_prob}
        \frac{w_{u\to v}^{\tau}}
        {1+\sum_{v'\in\mathcal N_t(u)}w_{u\to v'}^{\tau}}.
    \end{align}
    \State Set $i_{t+1}=v$, $\mathcal V_{\mathrm{prev}}\leftarrow
    \mathcal V_{\mathrm{prev}}\cup\{u\}$, $u=v$.
\EndFor
\State \Return $(\xv_{i_1},\xv_{i_T})$.
\end{algorithmic}
\end{algorithm}

\section{Graph-based adaptive importance sampling} \label{sec:PDD_estimation_IS}
In this section, we propose our graph-based adaptive importance sampling for PDD estimation. Our goal is to design the twisted sampling distribution $Q_X$ on the pairwise distance $r$, such that, the estimator $\widehat F_{\mathrm{IS}}(r)$ is accurate and sample efficient when the event $\{\rho(\xv_{2k-1},\xv_{2k}) \le r\}$ is rare. For different range of $r$, we propose an adaptive way for twisting $Q_X$.
Given a dataset of $N$ data samples, and the matrix of all $N^2$ distortions between pairs, it would be possible to construct a number of different twisted distribution that emphasize small distortions.
Our goal, however, was to develop a method with better than quadratic scaling in $N$ to allow analysis on very large datasets. Our graph-based approach has the following properties: i) The graph construction, done once per dataset, efficiently summarizes distortion information without requiring all $N^2$ distortions to be evaluated. ii) Selection of a pair of data points is done via a random walk on the graph - a simple operation. iii) Using an acyclic graph, there is only one path between any pair of nodes, which facilitates computing the twisted probability of any pair of points generated. 


\subsection{Non-backtracking random walk on MST}

In the pre-processing phase, we take a dataset of $N$ samples and construct an approximate minimum spanning tree of the of graph of all pairwise distortions. While an exact MST can be done in $O(N^2)$ operations, dominated by calculation of all distortions, an approximate MST can be calculated with lower complexity~\citep{zhong2013amst}. In our implementation, we first compute a k-nearest neighbor graph 
~\citep{friedman1977algorithm, andoni2015optimal}, and then compute an exact MST of the k-NN graph. Other methods of obtaining an approximate MST could also be used.
The next step is to give each directed edge in the tree a weight proportional to the number of nodes reachable via that edge. These weights can be used to generate a random walk that chooses pairs of nodes uniformly on the graph, or a twisted random walk that preferentially chooses short walks on the graph.
Details of the weight computation and random walk, as a function of a twist parameter $\tau$, are given in
Algorithm~\ref{alg:graph_sampler}.
The starting node is uniformly picked from $V$, and the random walk from the starting node is non-backtracking. It stops when it chooses itself as the next node, where we enforce the first node to not pick itself. The final sampled pair consists of the starting and ending nodes. The probability of picking the pair $(\xv_1, \xv_T)$ is the probability of taking the path $\xv_1 \to \xv_T$, where $T$ is the random stopping time. The path probability is $P(\xv_1,\cdots,\xv_T) = P(\xv_1) P(\xv_2 | \xv_1) \cdots P(\xv_T | \xv_{T-1})$, where $P(\xv_1) = 1/N$. When $\tau=1$ in the transition probability~\eqref{eq:transition_prob} (the untwisted case), the product of the conditional probabilities always turns out to be $P(\xv_2 | \xv_1) \cdots P(\xv_T | \xv_{T-1}) = 1/(N-1)$, so that the probability of selecting the pair is always $\mathbb{P}((\xv_1,\xv_T)) = 1/N(N-1)$. When $\tau \to 0$,  the transition probability at each node is more uniform, so the random walk becomes more likely to stop within fewer hops away from the starting node; in this way the probability of selecting nearby points is increased. By varying the parameter $\tau$, the twisted distribution that the pair is sampled from, $Q_{\tau}(\xv_1, \xv_T) = \mathbb{P}\big((\xv_1,\xv_T)\big) \neq 1/N(N-1)$, changes, controlling the  frequency with which nearby pairs are selected.
For different ranges of $r$,  we can adapt $\tau$ to keep $\mathrm{CV}\big(\widehat F_{\mathrm{IS}}(r)\big)$ small. To estimate $F(r)$ for small $r$, we choose smaller $\tau$ such that closer pairs are sampled. The correction ratio in the estimator is then $Y=(1/N(N-1))/\mathbb{P}(\xv_1, \xv_2, \cdots, \xv_T)$, where $(\xv_1, \xv_2, \cdots, \xv_T)$ is the actual random walk path.

\begin{figure*}[t]
    \centering
    \begin{subfigure}{0.24\linewidth}
        \centering
        \includegraphics[width=\linewidth]{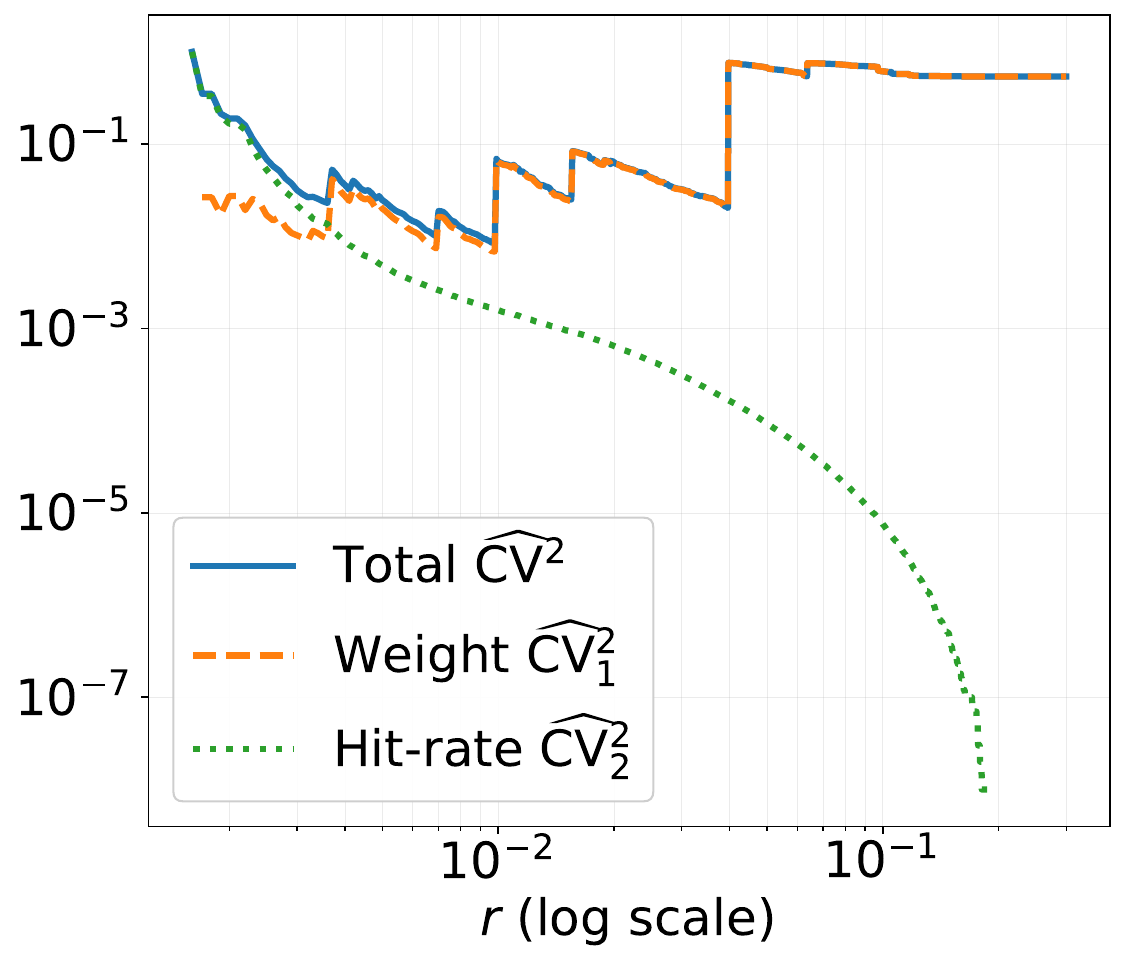}
        \caption{$\tau=0.2$}
    \end{subfigure}\hfill
    \begin{subfigure}{0.24\linewidth}
        \centering
        \includegraphics[width=\linewidth]{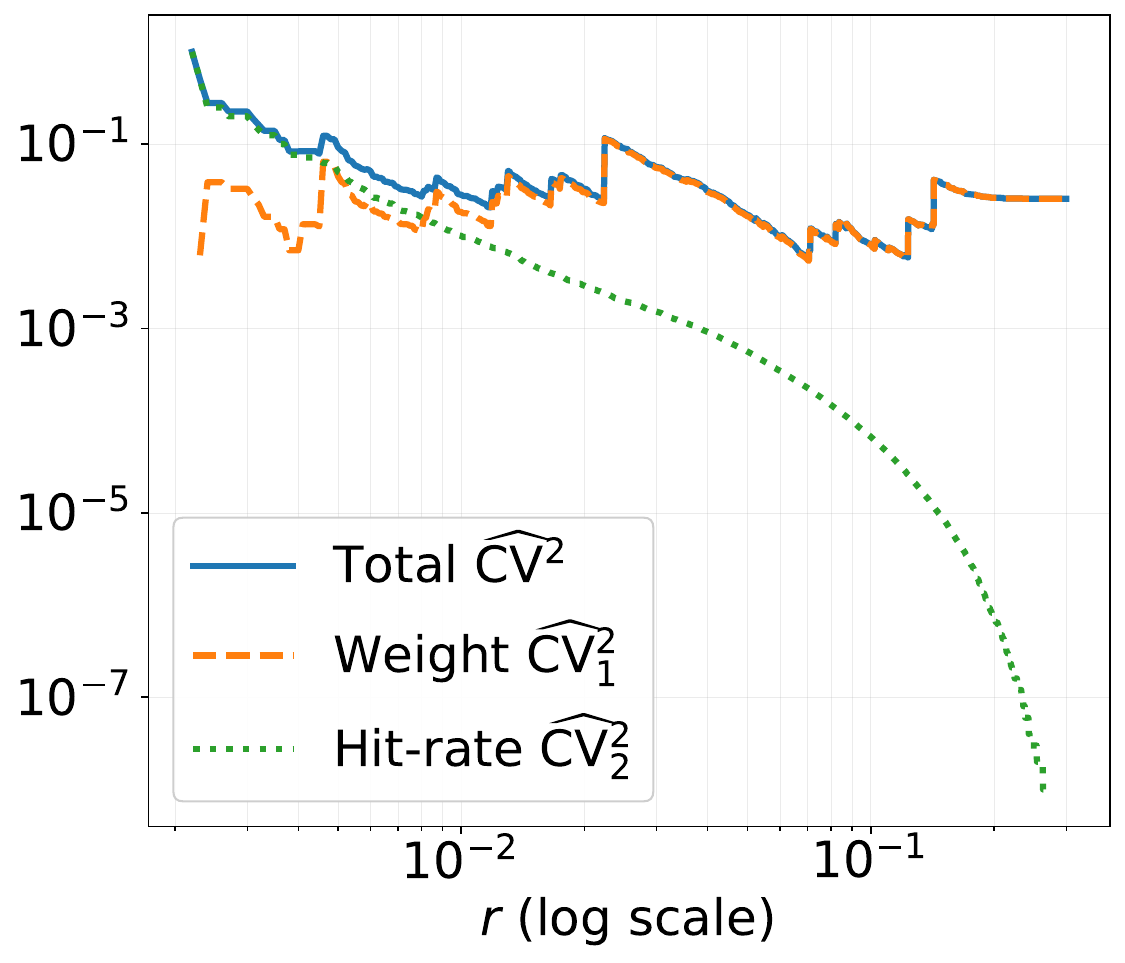}
        \caption{$\tau=0.5$}
    \end{subfigure}\hfill
    \begin{subfigure}{0.24\linewidth}
        \centering
        \includegraphics[width=\linewidth]{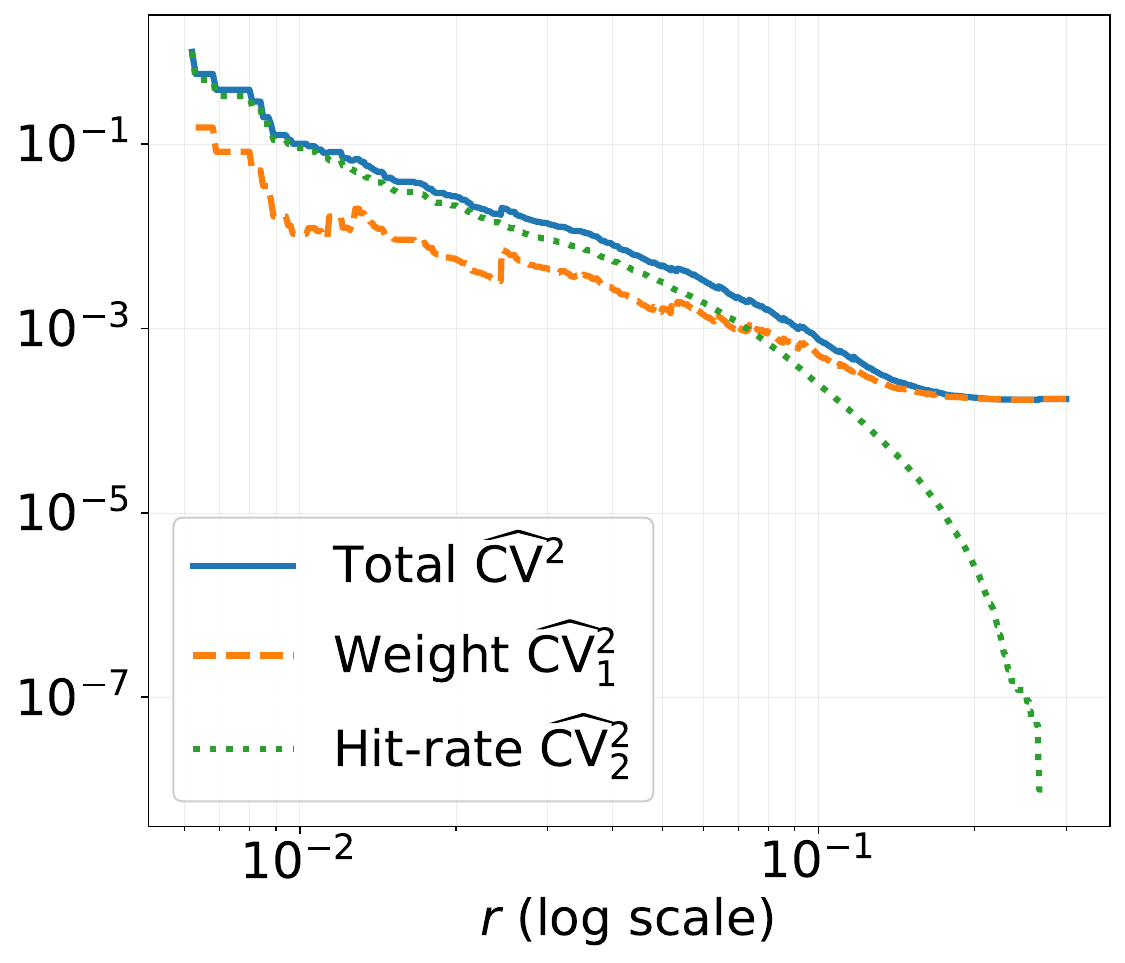}
        \caption{$\tau=0.8$}
    \end{subfigure}\hfill
    \begin{subfigure}{0.24\linewidth}
        \centering
        \includegraphics[width=\linewidth]{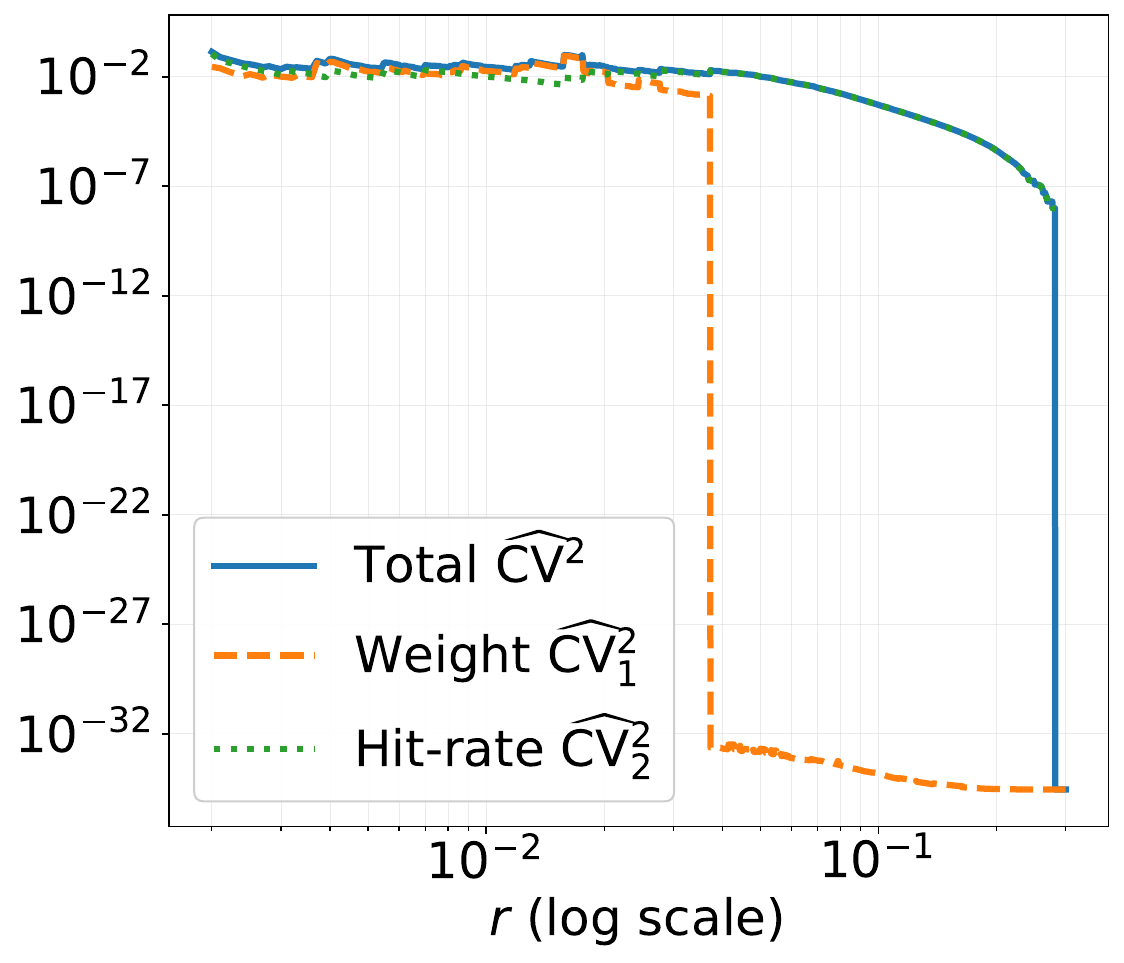}
        \caption{Adaptive $\tau$}
    \end{subfigure}
    \vspace{-2mm}
    \caption{The coefficient of variation of
    $\widehat{F}_{\text{IS}}(r)$ using different twisted distributions.}
    \label{fig:CV_tau}
    \vspace{-6mm}
\end{figure*}

\subsection{Variance analysis of the estimator} 

In Section~\ref{subsec:accuracy_of_IS}, we showed that the coefficient of variation of our estimate of $F(r)$ has two terms. The first term relates to the variability of the twisting ratio, conditioned on $\rho(x_1,x_2)\leq r$, and the second term is 
inversely proportional to the "hit-rate", or twisted probability that $\rho(x_1,x_2)\leq r$.
In the Figure~\ref{fig:CV_tau} (a), we plot an estimate of these two terms, and the total coefficient variation as a function of distance $r$ for the case that twisting parameter is $\tau = 0.2$.
The twisting ratio term (dashed line) tends to grow with $r$, while the hit-rate term (dotted line) tends to decrease with $r$. This twisted distribution is well-suited for estimating $F(r)$ in the vicinity of $r\approx 0.01$. The (b) and (c) figures show other that larger (less aggressive) twist parameters are better suited for estimating $F(r)$ for larger values of $r$. 

The CV estimates curves for the twisting ratio term (dashed lines) have large discrete jumps, particularly when $r$ is large and $\tau$ is small. This is evidence that this term can become heavy tailed if $\tau$ and $r$ are not well-matched.
In order to accurately and efficiently estimate $F(r)$ across a wide range of $r$, it is necessary to adapt $\tau$ as a function of $r$.

\subsection{Adaptive schedule of $\tau$}
Inspired by the observation above, we developed a strategy that gradually decreases $\tau$ with $r$. We begin with the untwisted distribution $\tau_0 = 1$ which is always suitable for estimating large values of $F(r)$.
We set an empirical threshold, say $\lambda = 0.1$, and find the largest value $r_0$ such $\widehat{\mathrm{CV}}\left(\hat{F}(r)\right) \geq \lambda$. We use $\tau_0$ to estimate $F(r)$ for $r\geq r_0$. We then reduce $\tau$, say to $\tau_1 = \tau_0-0.1$, and similarly compute $r_1$ such that 
$\widehat{\mathrm{CV}}\left(\hat{F}(r)\right) \geq \lambda$ for $r_1 \leq r \leq  r_0$. We use $\tau_1$ to estimate $F(r)$ in this interval. Continuing in this way, we obtain estimates of $F(r)$ with good accuracy across a range of scales. The Figure~\ref{fig:CV_tau} (d) shows the estimated coefficient of variation obtained for estimating $F(r)$, which stays below 0.01 over a wide range. Further details about the parameters $\tau_i$ and $r_i$ obtained in this example are given in Appendix~\ref{app:exp_tau}.
In the next section, we give numerical examples of estimating the PDD, along with RCQ bounds and PDD dimensions.

\section{Numerical results} \label{sec:Simulation}
In this section, we evaluate the RCQ bounds and the dimension for Gaussian source and MNIST. The details of calculating the PPD-based RCQ bounds, and the PDD dimension are in Appendix~\ref{app:PDD_RCQ_calculation} and \ref{app:PDD_dimension_calculation}. We compare our graph-based adaptive importance sampling (GAIS) against the MC uniform sampling.

\begin{figure*}[t]
    \centering
    \begin{subfigure}{0.24\textwidth}
        \centering
        \includegraphics[width=\linewidth]{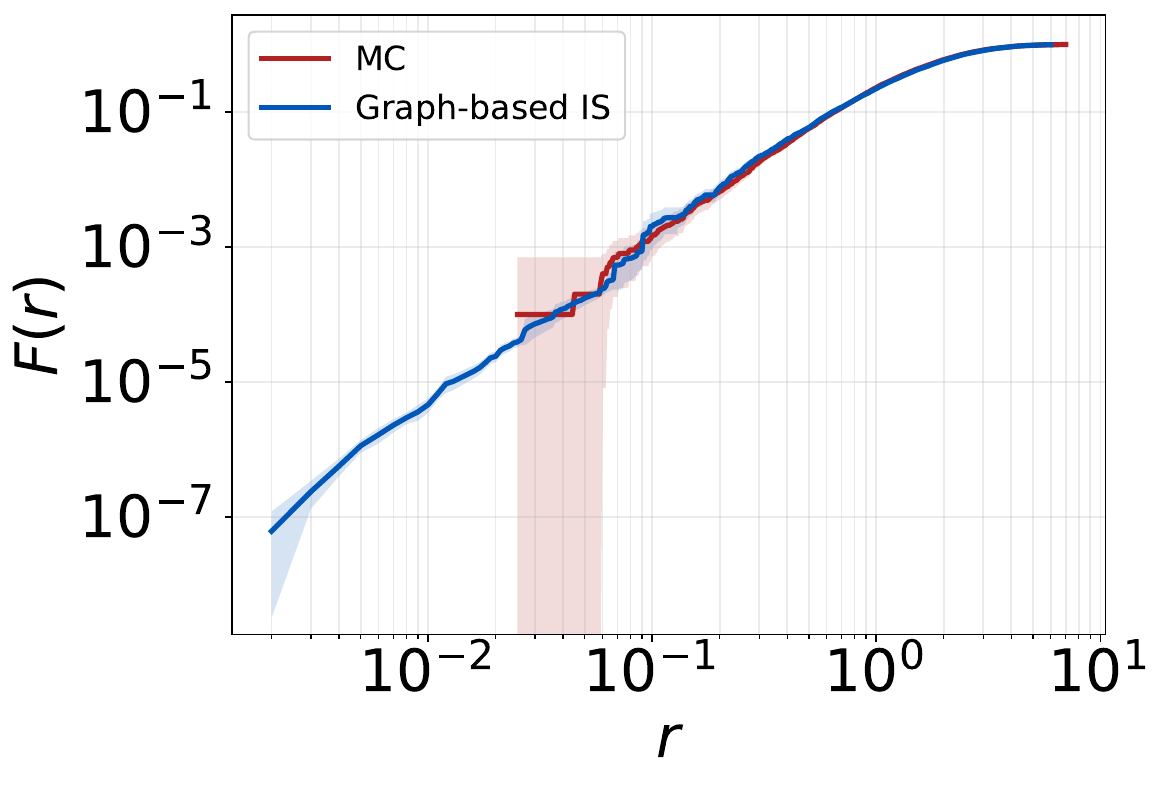}
    \end{subfigure}
    \hfill
    \begin{subfigure}{0.24\textwidth}
        \centering
        \includegraphics[width=\linewidth]{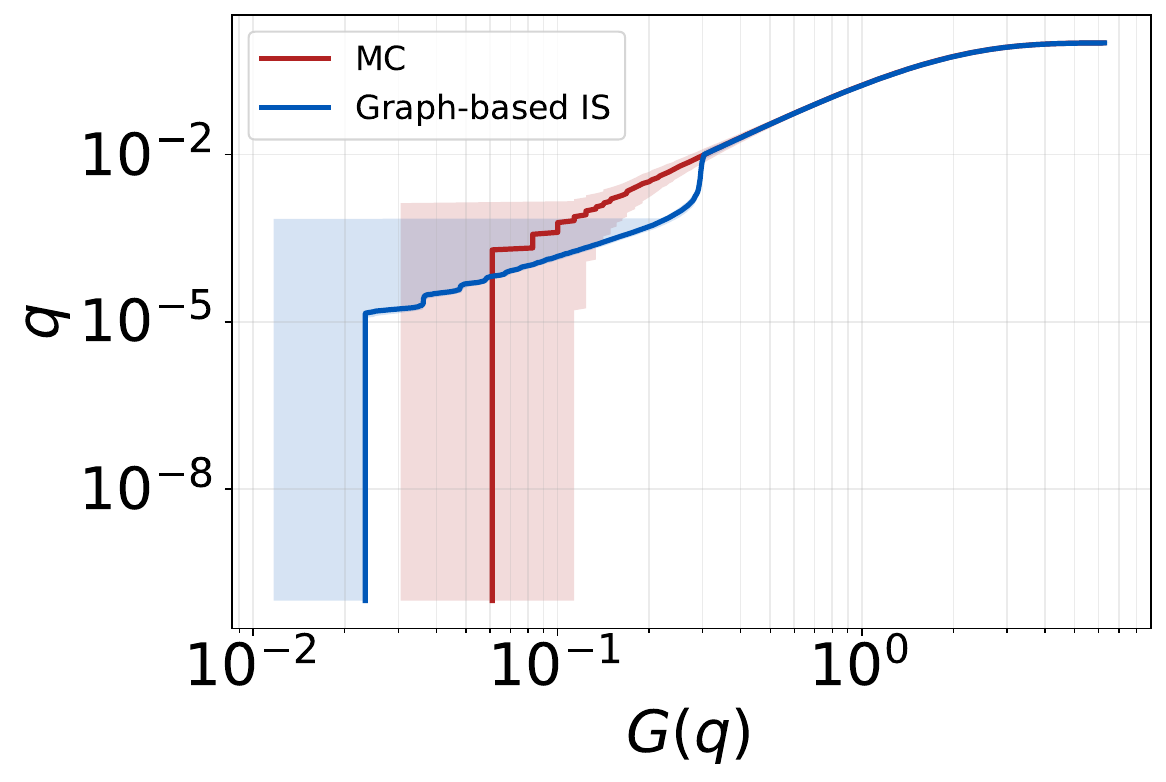}
    \end{subfigure}
    \hfill
    \begin{subfigure}{0.24\textwidth}
        \centering
        \includegraphics[width=\linewidth]{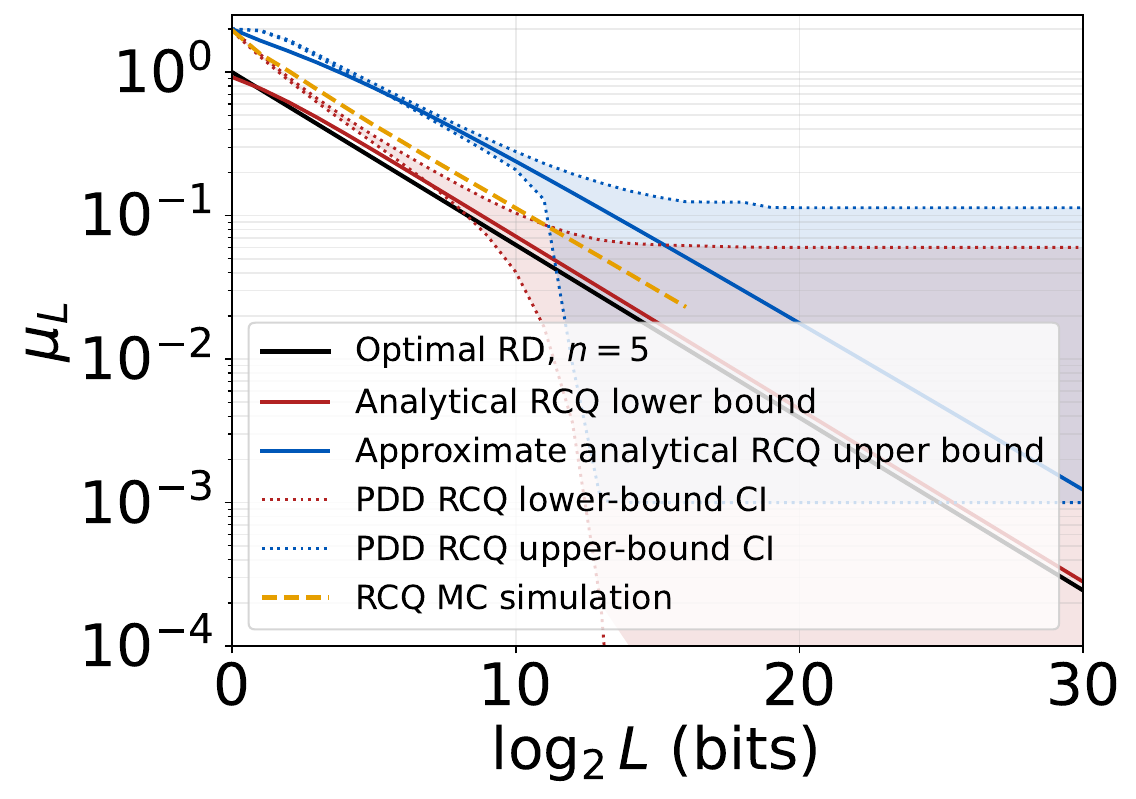}
    \end{subfigure}
    \hfill
    \begin{subfigure}{0.24\textwidth}
        \centering
        \includegraphics[width=\linewidth]{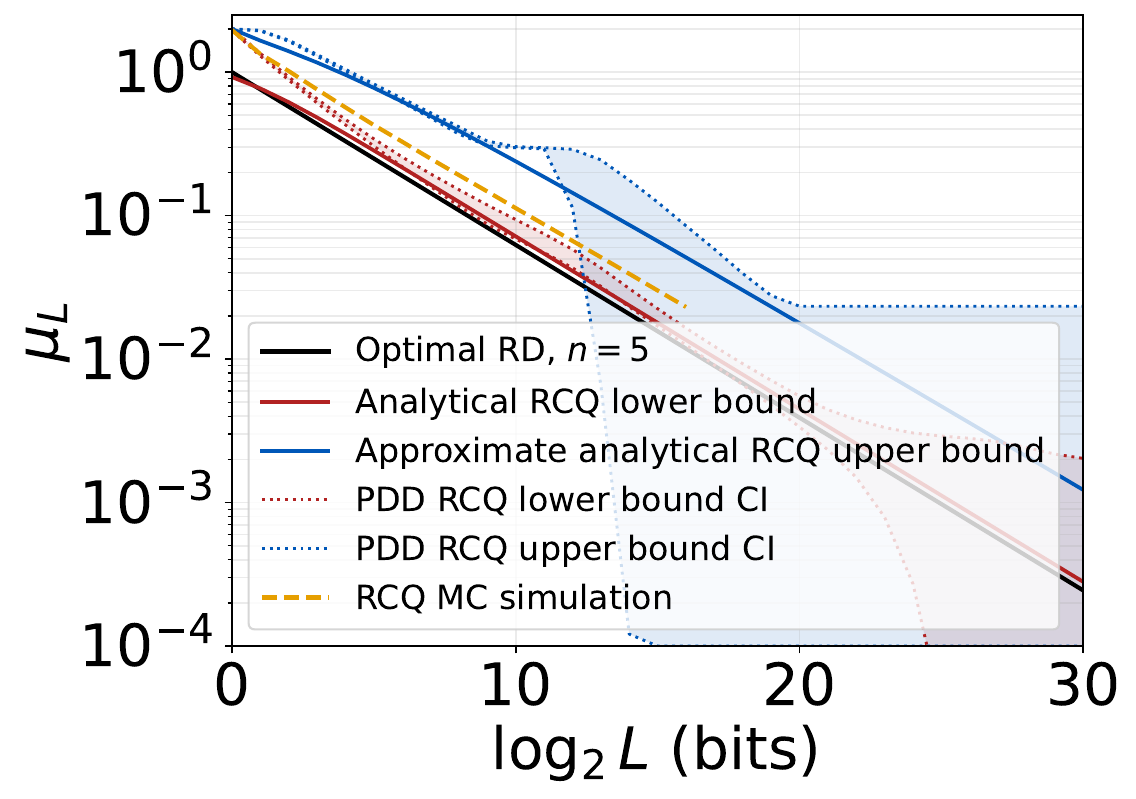}
    \end{subfigure}

    \begin{subfigure}{0.24\textwidth}
        \centering
        \includegraphics[width=\linewidth]{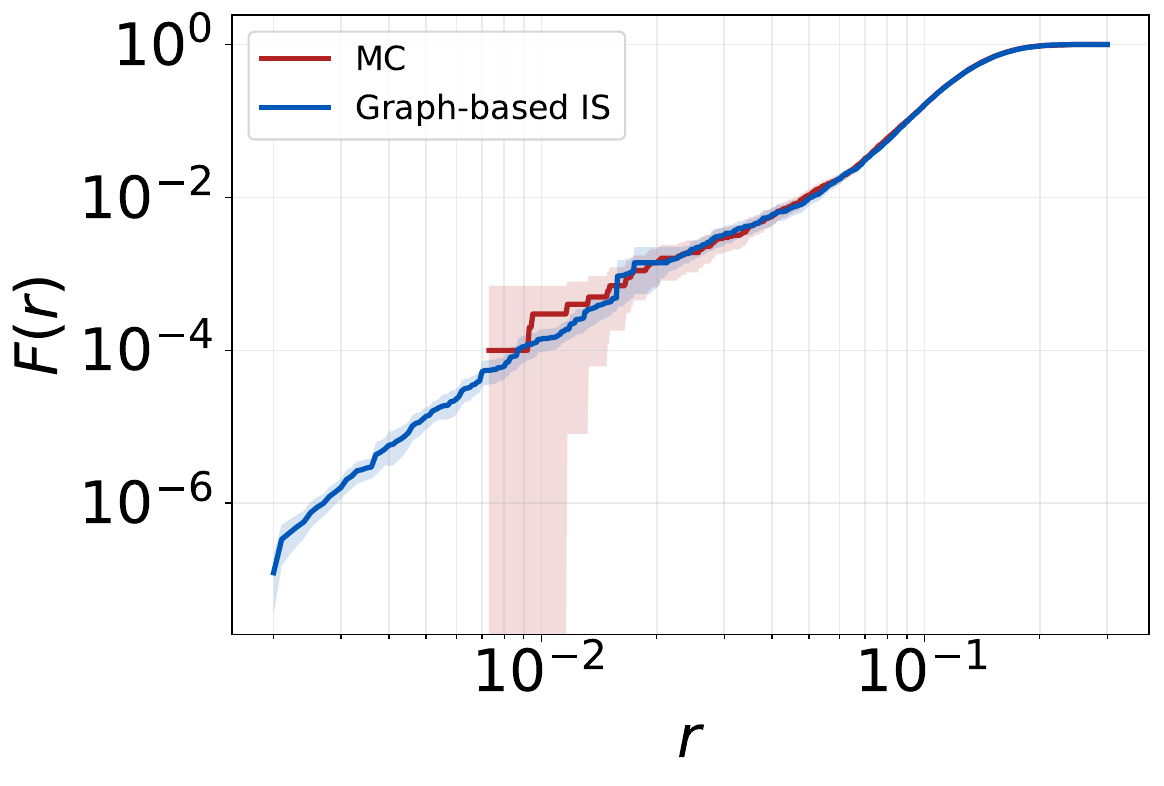}
    \end{subfigure}
    \hfill
    \begin{subfigure}{0.24\textwidth}
        \centering
        \includegraphics[width=\linewidth]{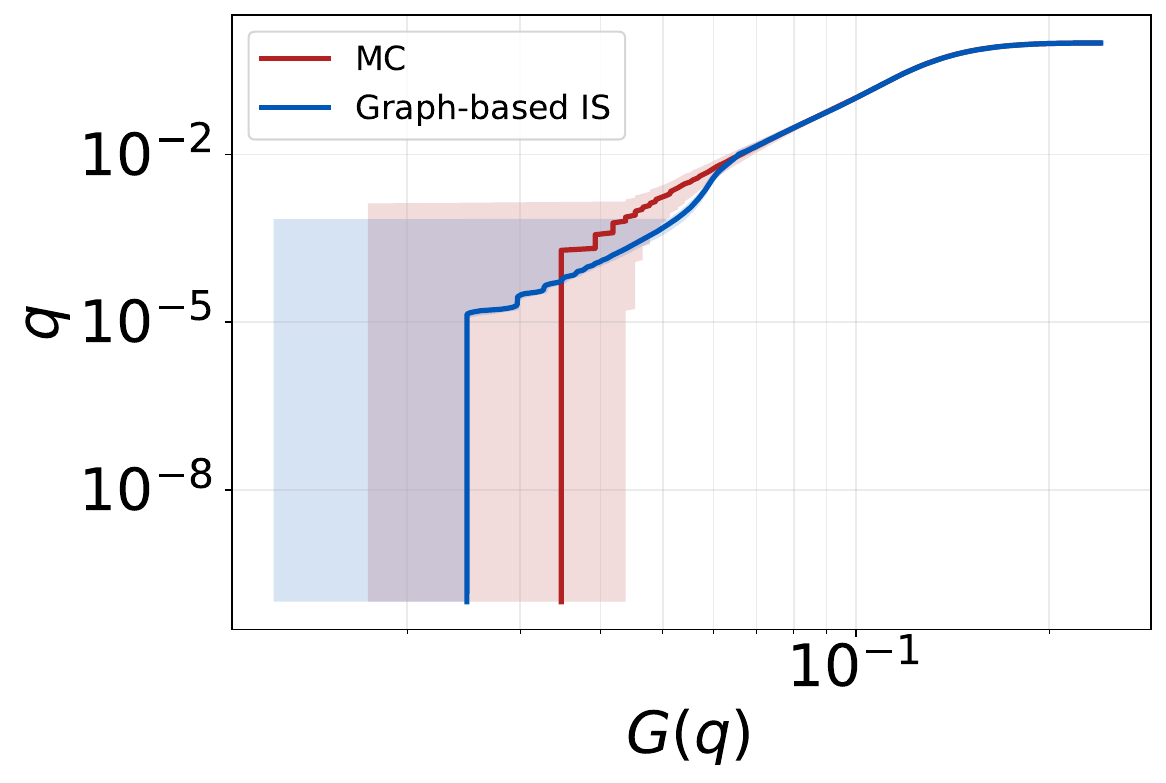}
    \end{subfigure}
    \hfill
    \begin{subfigure}{0.24\textwidth}
        \centering
        \includegraphics[width=\linewidth]{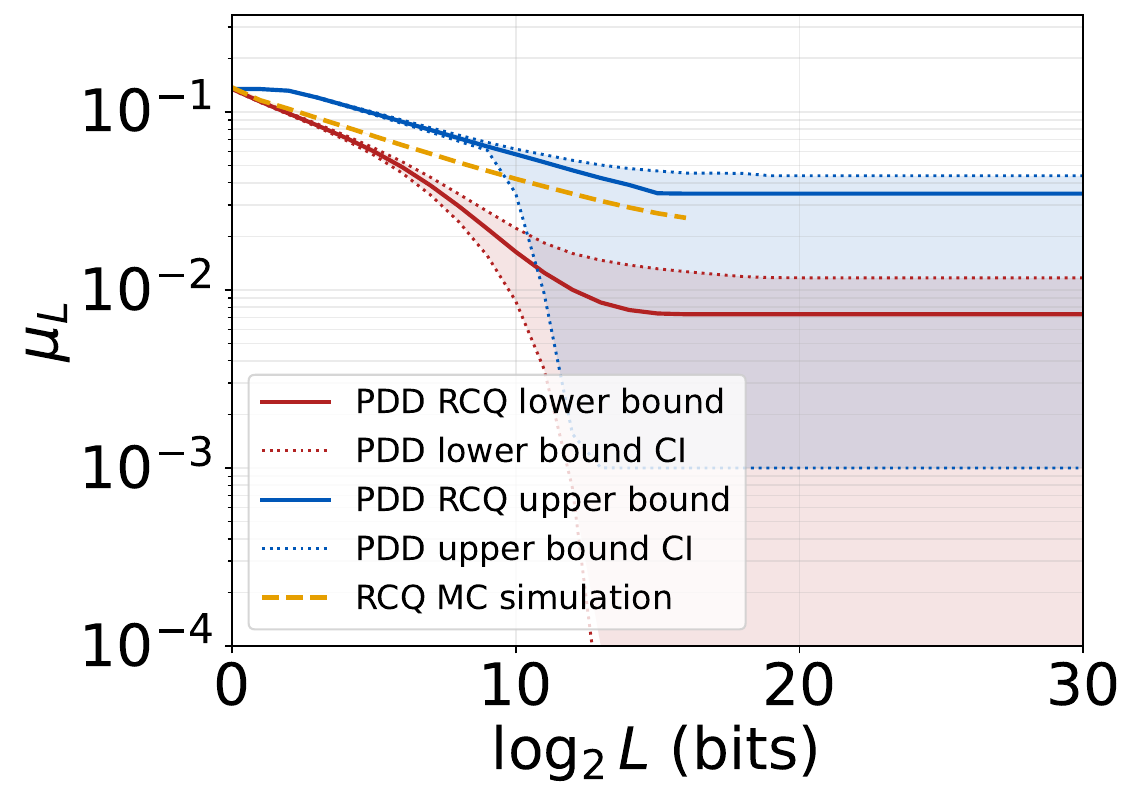}
    \end{subfigure}
    \hfill
    \begin{subfigure}{0.24\textwidth}
        \centering
        \includegraphics[width=\linewidth]{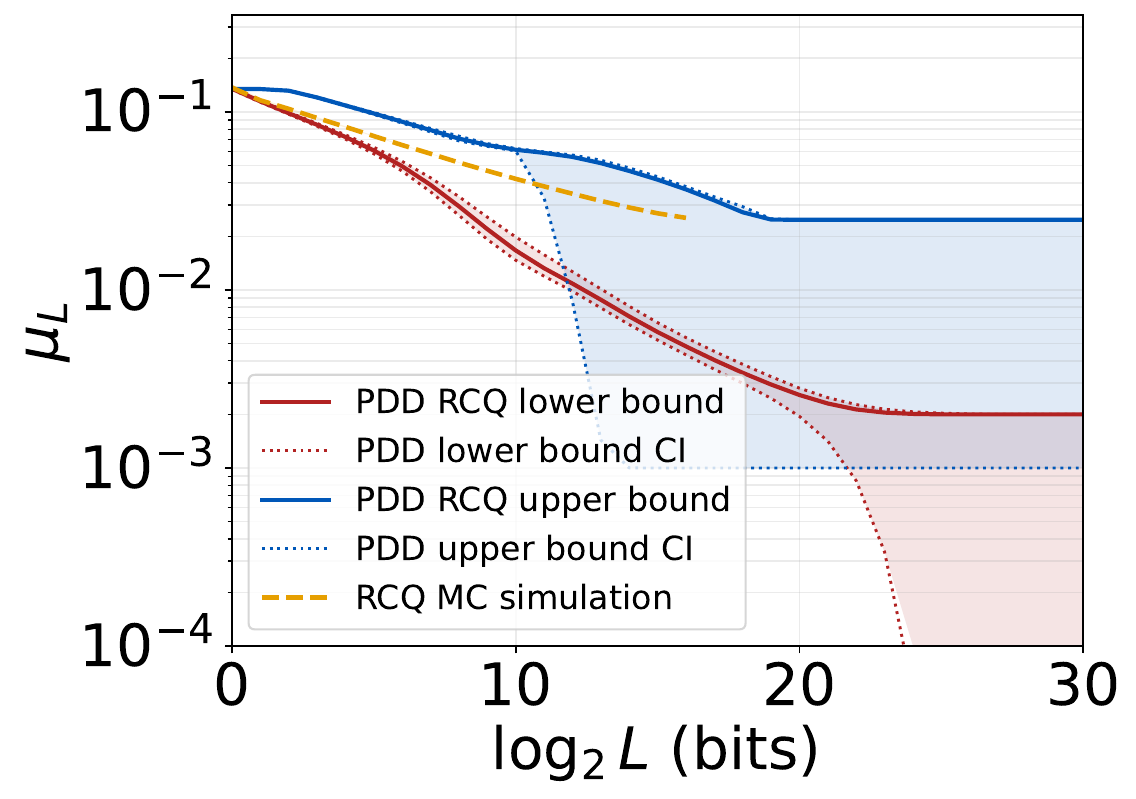}
    \end{subfigure}
    
    \vspace{-3.0mm}
    \caption{$F(r)$, $G(q)$, RCQ bounds estimated by MC uniform sampling, and GAIS (first row: multivariate Gaussian; second row: MNIST). We plot the confidence interval of estimating $F(r)$, $G(q)$, and the related RCQ bounds. For Gaussian source, we compute the optimal RD and the analytical RCQ lower and upper bounds.}
    \label{fig:PDD_F_G_RCQ}
    \vspace{-5mm}
\end{figure*}

\subsection{Multivariate Gaussian}
We first consider verify the methodology on a known distribution, $\xv \sim \mathcal{N}(\mathbf{0}, \Sigma)$, where $\Sigma = I_n$. We have the analytical RCQ lower bound based on Corollary~\ref{cor:F_bound}, and the approximate analytical upper bound based on Corollary~\ref{cor:G_bound} by assuming $\gamma = \beta^{-1/n}$, with $\nu=2$, $p=2$ in the distortion function, we have $\alpha=n/2$, $\beta=(4 \pi)^{-n/2} \text{det}(\Sigma)^{-1/2} \frac{\pi^{n/2}}{\Gamma(n/2 + 1)} r^{n/2}$. We use them as good references to evaluate estimation results by our proposed GAIS. The PDD estimation is performed on a finite set consisting of $60$K samples from the distribution. The number of samples used for the estimators is $10$K. In Figure~\ref{fig:PDD_F_G_RCQ} first row, we show that, compared with the MC sampling, $F(r)$ and $G(q)$ estimated using GAIS has a smaller confidence interval, and can estimate small probability region. The PDD-based RCQ bounds calculated using GAIS give a more stable and accurate bound especially in the high bitrate regime. We can find that the RCQ bounds estimated using uniform sampling saturates after around 10 bits. Our approach instead, can accurately characterize the bound until the regime of over 20 bits. This is due to improved accuracy of $\hat{F}(r)$ estimation. Since large $F(r)$ yields small $1-F(r)$, which contributes less to the RCQ lower bound approximation when $L$ is larger, by first-order Taylor approximation, $\left( 1 - F(r) \right)^L \approx e^{-LF(r)}$, this indicates that, we need to accurately estimate $F(r)$ to the probability level of $O(L^{-1})$. For instance, when $\log_2 L = 20$, $F(r)$ around $10^{-6}$ needs to be estimated. However, $\widehat{F}(r)$ estimated by MC uniform sampling is inaccurate at $10^{-4}$. In contrast, $\hat{F}_{\mathrm{IS}}(r)$ offers a low variance estimate till the probability level of $10^{-7}$, which thus yields a more accurate PDD-based RCQ bounds estimation in a higher bitrate regime. Further, estimating the slopes of a log-log plot of $F(r)$ and $G(q)$ yield estimates of PDD dimension. Figure~\ref{fig:PDD_dimension} in Appendix~\ref{app:exp_dimension} shows that PDD dimension estimated using GAIS approaches the true dimension when $r$ is small.

\subsection{MNIST}

To test our methodology on an unknown source distribution, we consider images from the MNIST data set. Using the MNIST training set as the data source, which contains $60$K images, and are vectorized. The number of samples used for the estimators is $10$K. Results are plotted in the second row of  Figure~\ref{fig:PDD_F_G_RCQ}. Since we do not have the analytical RCQ bounds, the accurately estimated PDD-based RCQ bounds provide valuable information of the limits of rate-distortion performance for compressing or quantizing this distribution. We can find that GAIS provides a stable PDD-based RCQ lower bound until over $20$ bits, while the MC uniform sampling estimator becomes inaccurate from $10$ bits.


\section{Conclusion}
In this paper, we 
derived lower and upper bounds for random codebook quantization based on the pairwise distortion distribution of the source. We also define notions of intrinsic dimension associated with the asymptotic behavior of the RCQ bounds. The dimensions we defined provide lower and upper bounds to rate-distortion dimension,  R\'enyi information dimension, and quantization dimension. We further propose a graph-based adaptive importance sampling approach for better PDD estimation, based on which, we can accurately characterize the RCQ bounds and its associated dimension of known and unknown source distributions with significantly reduced sample complexity.

\section{Detailed Proofs}\label{appendix:proofs}
In this section, we present proofs of results given in the main body of the paper.

\subsection{Proof of Theorem~\ref{theorem:1}: RCQ lower and upper bounds}
The RCQ error is lower bounded as
\begin{align*}
\mu_L(P_X, P_C) &= \mathbb{E}_X \mathbb{E}_C \Big[ \min_{1\leq \ell \leq L} \rho(\xv, \cv_\ell) \Big] \\
&= \mathbb{E}_X \bigg[ \mathbb{E}_{C}\Big[\min_{1\leq \ell \leq L} \rho(\xv, \cv_\ell)\Big] \bigg] \\
&= \mathbb{E}_X \bigg[ \int_0^\infty \Big(1 -\mathbb{P}\big(\rho(\cv;\xv)\le r\big) \Big)^L dr \bigg] \\
&= \int_0^\infty \mathbb{E}_X \bigg[ \Big(1 -\mathbb{P}\big(\rho(\cv;\xv)\le r\big) \Big)^L \bigg] dr \\
& \stackrel{(a)}{\geq} \int_0^\infty \bigg(1 -\mathbb{E}_X \Big[\mathbb{P}\big(\rho(\cv;\xv)\le r\big) \Big] \bigg)^L dr \\
& = \int_0^\infty \bigg(1 - \mathbb{P}_{X\sim P_X, C \sim P_C}\big(\rho(\xv,\cv)\le r\big) \bigg)^L dr \\
& \stackrel{(b)}{=} \int_0^\infty \Big(1 - F(r) \Big)^L dr,
\end{align*}
$(a)$ applies Jensens' inequality, $(b)$ assumes $P_X = P_C$. When $P_X = P_C$, denote $\mu_Q(L) = \mu_L(P_X, P_C)$, $\mathbb{P}_{\xv \sim P_X, \cv \sim P_C}\big(\rho(\xv,\cv)\le r\big) = \mathbb{P}_{\xv_1,\xv_2 \sim P_X}\big(\rho(\xv_1,\xv_2)\le r\big)$. The pairwise distortion distribution in Definition~\ref{def:1}.
Now we prove the upper bound of RCQ. For function $h(x) = x^L$, where $x \in [0,1]$, $L \in \mathbb{Z}_+$, we have
\begin{align}\label{eq:xtoL_upper}
    h(x) \leq t^{L-1}x + (1-t^L) \mathbbm{1}\{x \geq t\}, \quad t \in [0,1].
\end{align}
Then the upper bound of RCQ is
\begin{align*}
    \mu_Q(L) &= \mathbb{E}_X \bigg[ \int_0^\infty \Big(1 -\mathbb{P}\big(\rho(\cv;\xv)\le r\big) \Big)^L dr \bigg] \\
    &= \mathbb{E}_X \bigg[ \int_0^\infty \Big(1 -F(r;\xv) \Big)^L dr \bigg] \\
    & \stackrel{(a)}{\leq} \mathbb{E}_X \bigg[ \int_0^\infty t^{L-1} \big(1-F(r;\xv)\big) + (1 - t^L) \mathbbm{1}\{1 - F(r;\xv) \geq t\}  dr \bigg] \\
    & = \mathbb{E}_X \bigg[ t^{L-1} \int_0^\infty \big(1-F(r;\xv)\big) dr + (1 - t^L) \int_0^\infty \mathbbm{1}\{1 - F(r;\xv) \geq t\}  dr \bigg] \\
    & = \mathbb{E}_X \bigg[ t^{L-1} \int_0^\infty \mathbb{P}\big( \rho(\cv;\xv)\ge r \big) dr + (1 - t^L) \int_0^\infty \mathbbm{1}\{F(r;\xv) \leq 1-t\}  dr \bigg] \\
    & \stackrel{(b)}{=} \mathbb{E}_X \bigg[ t^{L-1} \mathbb{E}_C\big[\rho(\cv;\xv)\big] + (1 - t^L) \int_0^{G(1-t;\xv)} dr \bigg] \\
    & = \mathbb{E}_X \bigg[ t^{L-1} \mathbb{E}_C\big[\rho(\cv;\xv)\big] + (1 - t^L) G(1-t;\xv) \bigg] \\
    & = t^{L-1} \mathbb{E}_{X,C}\big[\rho(\cv;\xv)\big] + (1 - t^L) \mathbb{E}_X \big[ G(1-t;\xv) \big] \\
    & = t^{L-1} \mu_Q(1) + (1 - t^L) G(1-t).
\end{align*}
$(a)$ is obtained by applying \eqref{eq:xtoL_upper}. $(b)$ uses Definition~\ref{def:2}.

\subsection{Proof of Lemma~\ref{lemma:pdd_lb}: asymptotics of RCQ lower bound}

First for simplicity consider $d>0$.
Choose $\epsilon$ with $0<\epsilon< d$.
There exists $\delta>0$ such that
$r^{d+\epsilon} \leq F(r) \leq r^{d-\epsilon}$ for $0\leq r \leq \delta$.
Note that
\begin{eqnarray*}
    \int_\delta^\infty \left(1-F(r)\right)^L dr & = & \int_\delta^\infty \left(1-F(r)\right) \left(1-F(r)\right)^{L-1} dr \\
    & \leq & (1-F(\delta))^{L-1} \int_0^\infty (1-F(r)) dr \\
    & \leq & (1-\delta^{d+\epsilon})^{L-1} \int_0^\infty (1-F(r)) dr.
\end{eqnarray*}
As this term decays exponentially in $L$, it doesn't contribute to the limit of $\log(1/L)/\log(\check{\mu}_O)$.
Now considering $r\leq \delta$, (with $\delta \leq 1$ wlog), we have
\begin{eqnarray*}
    \int_0^\delta \left(1-F(r)\right)^L dr & \leq & 
    \int_0^\delta \left(1-r^{d+\epsilon}\right)^L dr \\
    & \leq & \int_0^1 \left(1-r^{d+\epsilon}\right)^L dr \\
    & \leq & B\left(L+1,\frac{1}{d+\epsilon} \right).
\end{eqnarray*}
Since $B(x,y)=\frac{\Gamma(x)\Gamma(y)}{\Gamma(x+y)}$ and $\frac{\Gamma(x)}{\Gamma(x+y)} \sim x^{-y}$ as $x\to\infty$, (see e.g. \cite[\href{https://dlmf.nist.gov/5.11.E12}{(5.11.12)}]{NIST:DLMF}
)
we see that $\lim_{L\to\infty} \log(1/L)/\log(\check{\mu}_L) \leq d+\epsilon$.

To lower bound the limit, note that
$\int_\delta^\infty \left(1-F(r)\right)^L dr \geq 0$, and
\begin{eqnarray*}
    \int_0^\delta \left(1-F(r)\right)^L dr & \geq & 
    \int_0^\delta \left(1-r^{d-\epsilon}\right)^L dr\\
    &\geq& \int_0^1 \left(1-r^{d-\epsilon}\right)^L dr - \int_\delta^1 \left(1-r^{d-\epsilon}\right)^L dr \\
    & \geq & B\left(L+1,\frac{1}{d-\epsilon}\right) - \left(1-\delta^{d-\epsilon}\right)^L.
\end{eqnarray*}
Considering that the beta term behaves as $L^{-1/(d-\epsilon)}$ and the other term decays exponentially with $L$, we see that  $\lim_{L\to\infty} \log(1/L)/\log(\check{\mu}_L) \geq d-\epsilon$.
As the bounds hold for any $\epsilon > 0$, the desired limit is established.

In the case $d = 0$, the previous argument shows that the limit is less than or equal to $\epsilon$ for any $\epsilon>0$. That also establishes that $\check{\mu}_L \to 0$, so eventually the ratio $\log(1/L)/\log(\check{\mu}_L)$ is non-negative, so that its limit must also be non-negative.

\subsection{Proof of Lemma~\ref{lemma:pdd_ub}: asymptotics of RCQ upper bound}

{\emph Proof:}
It is convenient to denote 
$f_1(t)=t^{L-1}\mu_1$ and $f_2(t) =(1-t^L)G(1-t)$.
Rather than explicitly calculating the infimum of $f_1(t)+f_2(t)$, it is sufficient to work with upper and lower bounds on the infimum.
An upper bound is simply $f_1(t)+f_2(t)$ for any particular $t$.
For the lower bounds, note that for any $t'$
the infimum over $t\in [0,1]$ is equal to the minimum of the infimum over $t\in [0,t']$ and the infimum over $t\in [t',1]$. 
Because $f_2(t)$ is decreasing and $f_1(t)\geq 0$, 
$$
\inf_{0\leq t \leq t'} 
\left[ f_1(t)+f_2(t) \right] \geq f_2(t').
$$
Likewise, because $f_1(t)$ is increasing and $f_2(t)\geq 0$,
$$
\inf_{t' \leq t \leq 1} 
\left[ f_1(t)+f_2(t) \right] \geq f_1(t').
$$
Thus for any chosen $t\in[0,1]$, we have
$ \min{f_1(t),f_2(t)} \leq \hat{\mu}_L \leq 
f_1(t)+f_2(t)$
To obtain the desired limit, we will choose a sequence $t_L$ such $\lim_{L\to\infty} \frac{\log(1/L)}{\log(f_k(t_L))} = d$ for $k=1,2$. This then established the desired limit for $\hat{\mu}_L$.

Define $t_L = 1-\frac{\log(L)}{L d }$. Then 
\begin{eqnarray*}
\log\left(f_1(t_L)\right) & = &  (L-1)\log\left(1-\frac{\log(L)}{Ld}\right) + \log(\mu_1)\\
& \approx & -\frac{(L-1)\log(L)}{Ld} + \log(\mu_1).
\end{eqnarray*}
using the first order approximation of $\log$ near 0, from which the desired limit for $f_1$ follows.

As the calculations for $f_1$ show, $t^L$ goes to zero and so the term $(1-t_L^L)$ in $f_2(t_L)$ approaches unity in the limit. We thus need only focus on the behavior of $G(1-t_L)$.
\begin{eqnarray*}
    \lim_{L\to\infty} \frac{\log(1/L)}{\log(G(1-t_L)))}
    & = & \lim_{L\to\infty} \frac{\log(1/L)}{\log(G(\frac{\log(L)}{Ld})))} \\
    & = & \lim_{L\to\infty} \frac{\log\left(\frac{\log(L)}{Ld}\right)}{\log(G(\frac{\log(L)}{Ld})))} \frac{\log(1/L)}{\log\left(\frac{\log(L)}{Ld}\right)} \\
    & = & d \lim_{L\to\infty}\frac{\log(1/L)}{\log(1/L)+\log\left(\frac{\log(L)}{d}\right)} \\
    & = & d.
\end{eqnarray*}

\subsection{Proof of Corollary~\ref{cor:G_bound}: explicit bound on RCQ upper bound}

\emph{Proof:}
The result is obtained from Theorem 1 by noting
\begin{eqnarray*}
\overline{\mu}_Q(L) & = & 
\inf_{t\in [0,1]} \left[ t^{L-1} \mu_Q(1) + (1-t^L)G(1-t)\right] \\
& \leq & 
\inf_{t\in [0,1]} \left[ t^{L-1} \mu_Q(1) + G(1-t)\right] \\
& \leq & t_L^{L-1} \mu_Q(1) + G(1-t_L) \\
& \leq & t_L^{L-1} \mu_Q(1) + \beta \left(1-t_L\right)^{1/\alpha} \\
& \leq & \mu_Q(1) \left(1-s_L^{-1}\right)^{L-1} + \beta s_L^{-1/\alpha}.
\end{eqnarray*}
where we made the choice $s_L =(1-t_L)^{-1} = (L\alpha)/\log(L)$.
The purpose of this choice is so that both terms in the sum behave like $L^{-1/\alpha}$, up to a $\log(L)$ factor.

\subsection{Proof of Theorem~\ref{theorem:cts-case}: Asymptotic Bounds for Continuous Distributions}

\emph{Proof:}
Given $\epsilon>0$, choose $\delta>0$ such that $\rho(x,y)\leq \delta$ implies $|p(x)-p(y)|<\epsilon^2$.
Define $S(\epsilon) = \{ x\in\Real^N : p(x)\geq \epsilon\}$
For $x\in S(\epsilon)$ and $r <\delta^\nu$, we have
\begin{eqnarray*}
F(r;x) & = & \int_{\rho(x,y)\leq r} p(y) dy \\
& = & \int_{\|x-y\|_p \leq r^{1/\nu}} p(y) dy \\
& \leq & V_{p,N}(1) r^{N/\nu} \left(p(x)+\epsilon^2 \right) \\
& \leq & V_{p,N}(1) r^{N/\nu} p(x)(1+\epsilon).
\end{eqnarray*}
Similarly, $F(r;x) \geq V_{p,N}(1) r^{N/\nu} p(x)(1-\epsilon)  $. Multiplying these inequalities by $p(x)r^{-N/\nu}$, then integrating the result over $S(\epsilon)$, and finally taking the limit as $\epsilon\to 0$, gives the desired expression for $F(r)$.

For $x\in S(\epsilon)$, the bounds on $F(r;x)$ imply that
$$
\left(\frac{q}{V_{p,N}(1) p(x)(1+\epsilon)}\right)^{\nu/N} \leq G(q;x) \leq \left(\frac{q}{V_{p,N}(1) p(x)(1-\epsilon)}\right)^{\nu/N}.
$$
Again multiplying these inequalities by $p(x)q^{-\nu/N}$, integrating the result over $S(\epsilon)$, and then letting $\epsilon\to 0$ gives the desired expression for $G(q)$.

\subsection{Proof of Theorem~\ref{theorem:calibrated}: Connections between different notions of dimension}

It is helpful to first state and prove some results about uncalibrated dimensions.

\begin{lemma}\label{lemma:uncalibrated}
    For an arbitrary distortion measure, the uncalibrated dimensions satisfy
    $$
    \tilde{\underline{d}}_Q(X) \leq \tilde{d}_Q(X) \leq \tilde{\overline{d}}_Q(X),
    $$
    and 
    $$
    \tilde{d}_U(X) \leq \tilde{d}_O(X) \leq \tilde{d}_Q(X).
    $$
    whenever these quantities exist.
\end{lemma}
\emph{Proof:}
These follow from the relationships between corresponding distortion curves $\mu(L)$. In particular, consider two distortion curves $\mu_1(L) \geq \mu_2(L) > 0$, with corresponding dimensions
$$\lim_{L\to\infty} \frac{\log(L)}{\log(1/\mu_k(L))} = d_k > 0,
$$
for $k=1,2$. Then we can show $d_1 \geq d_2$. To this end, define $g(L) = \mu_2(L)/\mu_1(l)$. 
$$
\lim_{L\to\infty} \frac{\log(L)}{\log(1/g(L))} = \frac{1}{1/d_2 - 1/d_1}=\frac{d_1 d_2}{d_1-d_2}.
$$
Since $\log(1/g(L))\geq 0$, we must have $d_1 - d_2 \geq 0$. This concludes the proof of the lemma, for $\tilde{d}_U(X)$, $\tilde{d}_O(X)$, and $\tilde{d}_Q(X)$.  For $\tilde{\underline{d}}_Q(X)$ and $\tilde{\overline{d}}_Q(X)$, the proof follows after connecting the behavior of the RCQ bounds $\underline{\mu}_Q(X)$ and $\overline{\mu}_Q(X)$ to the corresponding PDD dimensions, via Lemmas~\ref{lemma:pdd_lb} and \ref{lemma:pdd_ub}.

We now proceed with the proof of Theorem~\ref{theorem:calibrated}.
Most of the relationships are obtained  directly from Lemma \ref{lemma:uncalibrated} upon multiplication through by the distortion exponent $\nu$. The identification of Renyi information dimension and rate-distortion dimension, $d_I(X) = d_U(X)$, for given distortion function is given in \cite{wu2010renyi}, example. It remains to show $\underline{d}_Q(X) \leq d_I(X)$.

Proposition 4 in \cite{wu2010renyi}
 shows that for a $p$-norm on $\Real^n$ with $1\leq p \leq \infty$,
 \begin{equation}\label{eq:rid_formula}
 \vspace{-0.02in}
 d_I(X) = \lim_{r\to 0} \frac{\Exp{X_1}{\log \Probd{X_2}{\norm{X_1-X_2}_p\leq r}}}{\log r}.
 \end{equation}
 For each $r$,
 \begin{eqnarray*}
E_{X_1} \left[ \log \Probd{X_2}{\norm{X_1-X_2}_p \leq r} \right]
 & = & 
\Exp{X_1}{\log \Exp{X_2}{1\left\{ \norm{X_1-X_2}_p\leq r\right\}}} \\
& \leq  &
\log \Exp{X_1}{ \Exp{X_2}{1\left\{ \norm{X_1-X_2}_p\leq r \right\}}} \\
& = &
\log \Probd{X_1, X_2}{\rho(X_1,X_2)\leq r^\nu}
\; = \; \log F(r^\nu).
\end{eqnarray*}

Dividing both sides by  $\log r^\nu$, which is negative for $r<1$, and taking the limit as $r\to 0$ then establishes that 
$$
d_I(X) \geq \nu \lim_{r\to 0} \frac{\log(F(r))}{\log(r)}.
$$





\bibliography{reference}
\bibliographystyle{plainnat}

\clearpage
\appendix

\onecolumn

\section{Additional analysis and experimental results}
We provide the additional experimental details and results in this section to better illustrate our proposed method and its performance.

\subsection{PDD-based RCQ bounds calculation}
\label{app:PDD_RCQ_calculation}
We first estimate the PDD of the pairs sampled from $P_X$ using the proposed adaptive graph-based importance sampling $\widehat F_{\mathrm{IS}}(r_1), \cdots, \widehat F_{\mathrm{IS}}(r_m)$, where $r_1 < r_2 \cdots < r_m$ cover almost the whole range. For estimation at each $r$, we denote the $\pm 95\%$ confidence interval (CI) of $\widehat F_{\mathrm{IS}}(r_m)$ as $\widehat{F}_{\overline{\text{CI}}}$ and $\widehat{F}_{\underline{\text{CI}}}$. So, to approximate the integral in the lower bound of $\mu_L$ in Theorem~\ref{theorem:1}, the empirical lower bound is within the range
\[
\left[\sum^{M-1}_{m=1} \big( 1 - \widehat{F}_{\overline{\text{CI}}}(r_{m+1}) \big)^L (r_{m+1} - r_m), \sum^{M-1}_{m=1} \big( 1 - \widehat{F}_{\underline{\text{CI}}}(r_m) \big)^L (r_{m+1} - r_m)\right].
\]
We set the $\widehat{F}_{\overline{\text{CI}}}(r)$ to be $7/K$ if no hits are observed in $K$ samples, which is a heuristic value giving a conservative CI. It is based on an argument that when observing $K$ Bernoulli random variables probability as high as $7/K$, there is a non-negligible probability of observing no hits.

Now we calculate the RCQ upper bound in Theorem~\ref{theorem:1} based on the PDD estimation. For a given probability $q$, we first randomly sample a $\xv$, and apply the graph-based importance sampling  $K$ times to obtain $\widehat{F}_{\text{IS}}(r;\xv)$. 
The corresponding inverse function is then estimated as
\[
\hat{G}_{IS}(q;\xv)=\max\{r:\hat{F}_{IS}(r;\xv)\leq q\},
\]
with confidence intervals on $\hat{G}$ likewise calculated from confidence intervals on $\hat{F}$.

This procedure is repeated for $M$ randomly sampled $\xv_i$, for each of them we obtain $\hat{G}_{IS}(q;\xv_i)$, then we take the average of $M$ of them to obtain the term
\[
\hat{G}_{IS}(q) = 
= \frac{1}{M} \sum^{M}_{i=1} \hat{G}_{IS}(q;\xv_i).
\]
The single codeword RCQ error in the upper bound is obtained as
\[
\hat{\mu}_Q(1) 
= \frac{1}{K_1} \sum^{K_1}_{k_1=1} \frac{1}{K_2} \sum^{K_2}_{k_2=1} \rho(\xv_{k_1}, \xv_{k_2}).
\]
For each value of $L$, the upper bound $\overline{\mu}_Q(L)$ is estimated by a numerical minimization of
$$
t^{L-1}\hat{\mu}_Q(1) + (1-t^L)\hat{G}_{IS}(1-t),
$$
for $t\in[0,1]$.
Confidence intervals for the bound are obtained by substituting confidence intervals for $\hat{G}$ into the minimization.

\subsection{PDD-based dimension calculation}
\label{app:PDD_dimension_calculation}
By applying l'Hopital's rule, the dimensions defined in Definition~\ref{def:3} are computed using estimated $\widehat{F}(r)$ and $\widehat{G}(q)$:
\[
\underline d_Q(X) \approx \lim_{r\to0} \frac{
\log \widehat F(r_{m+1}) - \log \widehat F(r_{m-1})}{\log r_{m+1} - \log r_{m-1}},
\]
and
\[
\overline d_Q(X) \approx \lim_{q\to0} \frac{\log q_{m+1} - \log q_{m-1}}{\log \widehat G(q_{m+1}) - \log \widehat G(q_{m-1})}.
\]
The ratios given above can be plotted as a function of $\log(r_m)$ or $\log(G(q_m))$; if the function tends to flat "plateau" in the small $r$ or small $q$ regime, then the level of this plateau  can be taken as an estimate of the corresponding dimension.
However, the values of the plotted ratios in the whole range of $r$ can be of interest in understanding the structure of a data source, which may effectively exhibit different characteristic dimensions at different scales. 

\subsection{Adaptive $\tau$ selection}
\label{app:exp_tau}
\begin{figure}[t]
    \centering
    \includegraphics[width=0.49\linewidth]{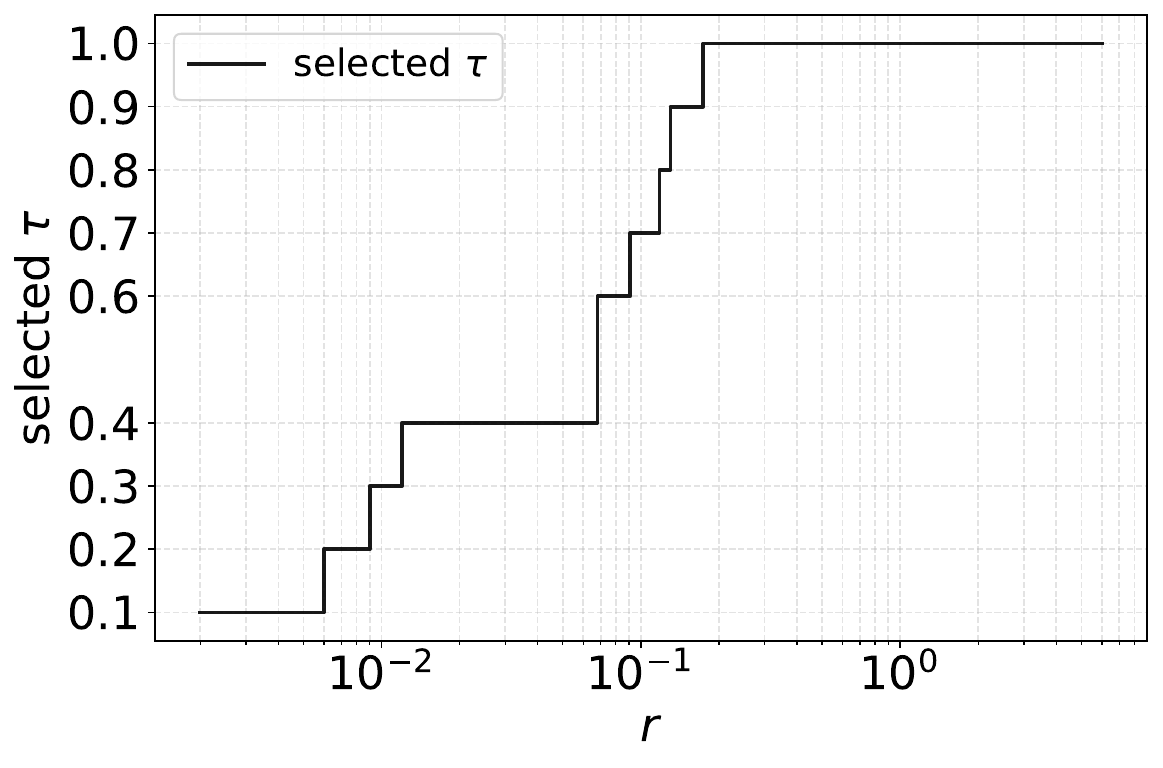}
    \hfill
    \includegraphics[width=0.49\linewidth]{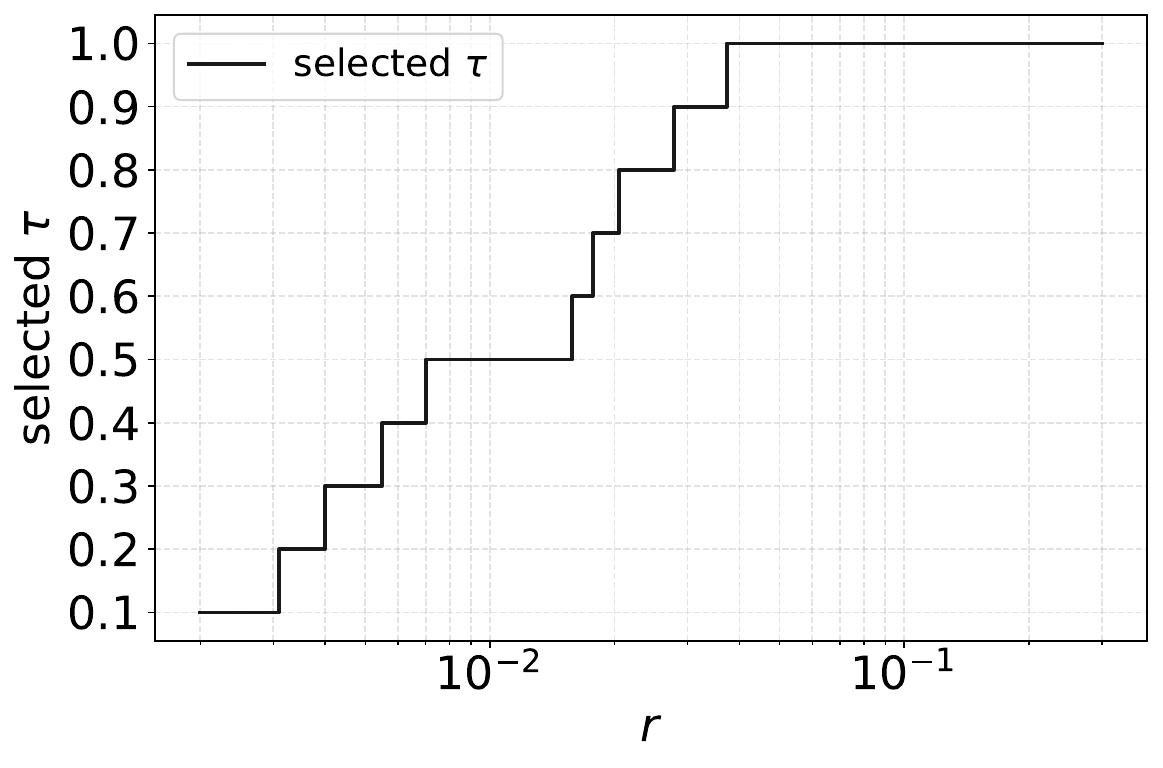}
    \caption{Adaptive $\tau$ choice in our proposed graph-based adapative importance sampling approach. Left: $5$-dimensional multivariate Gaussian. Right: MNIST. The $Q_\tau$ is twisted more (smaller $\tau$) when $r$ is smaller.}
    \label{fig:tau_choice}
\end{figure}

In Figure~\ref{fig:tau_choice}, we show the selected $\tau$ of our proposed Graph-based adaptive importance sampling for $5$-dimensional multivariate Gaussian source (left) and MNIST training set (right). The initial $\tau$ is $1.0$, the step size is $0.1$. The adaptive decreasing criterion for $\tau$ is
\[
\sum_{k=1}^{K} \mathbbm{1}\{\rho(\xv_1,\xv_T) \le r\} \ge 50, \quad \widehat{\mathrm{CV}(\hat{F}_{\mathrm{IS}}(r))^2} \leq 0.1.
\]
We can find that, at different regions of $r$, a suitable twisted distribution $Q_{\tau}$ controlled by the parameter $\tau$ is used for sampling the pair $(\xv_1, \xv_T)$.

\subsection{PDD dimension estimation}
\label{app:exp_dimension}
\begin{figure*}[t]
    \centering
    \begin{subfigure}{0.24\textwidth}
        \centering
        \includegraphics[width=\linewidth]{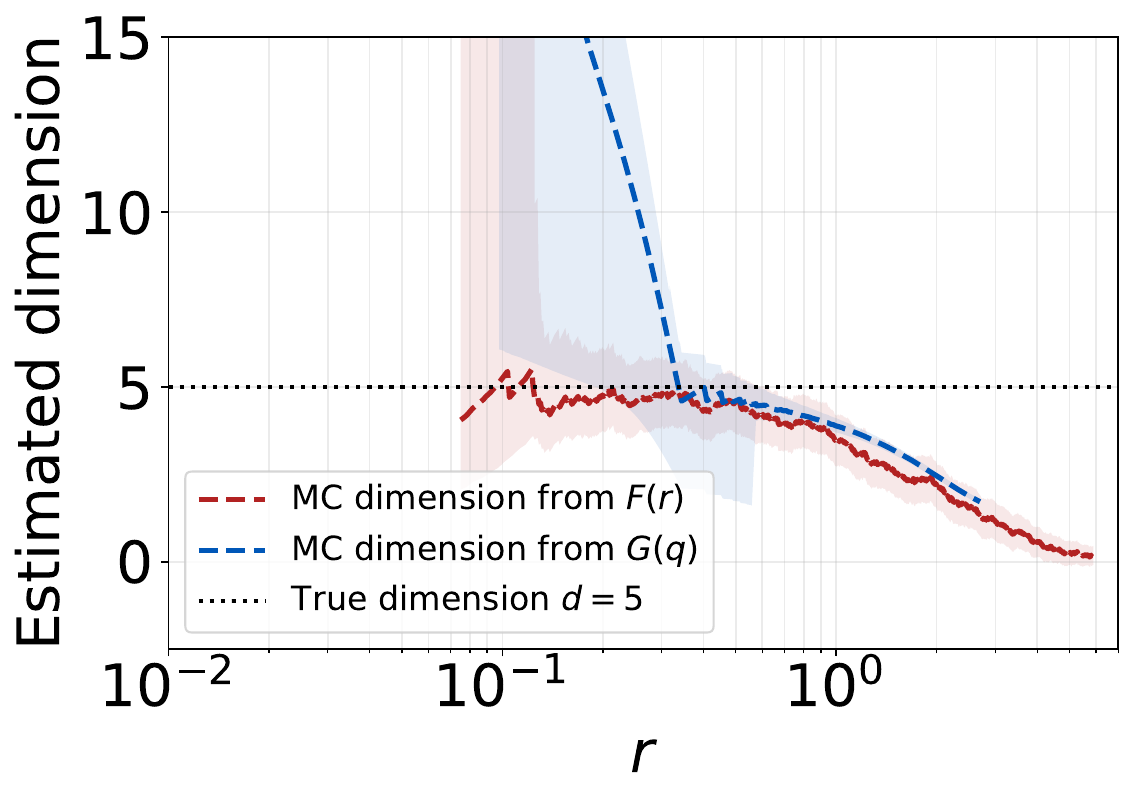}
    \end{subfigure}
    \hfill
    \begin{subfigure}{0.24\textwidth}
        \centering
        \includegraphics[width=\linewidth]{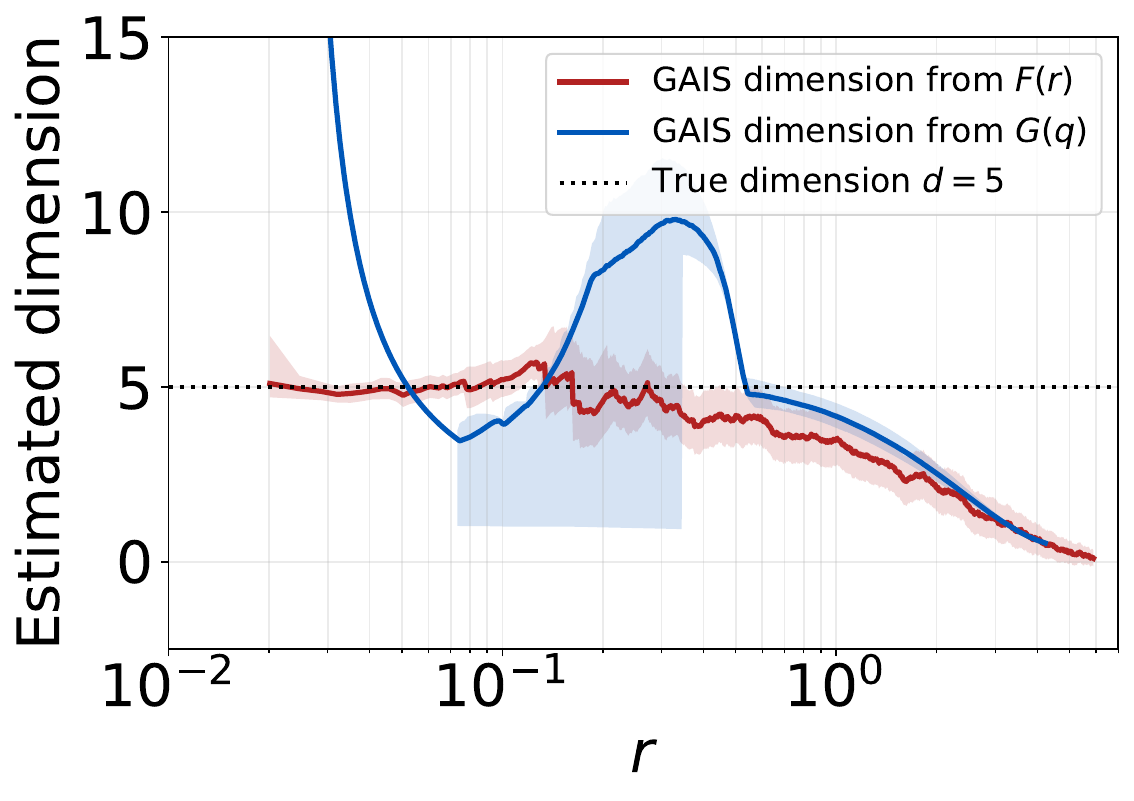}
    \end{subfigure}
    \hfill
    \begin{subfigure}{0.24\textwidth}
        \centering
        \includegraphics[width=\linewidth]{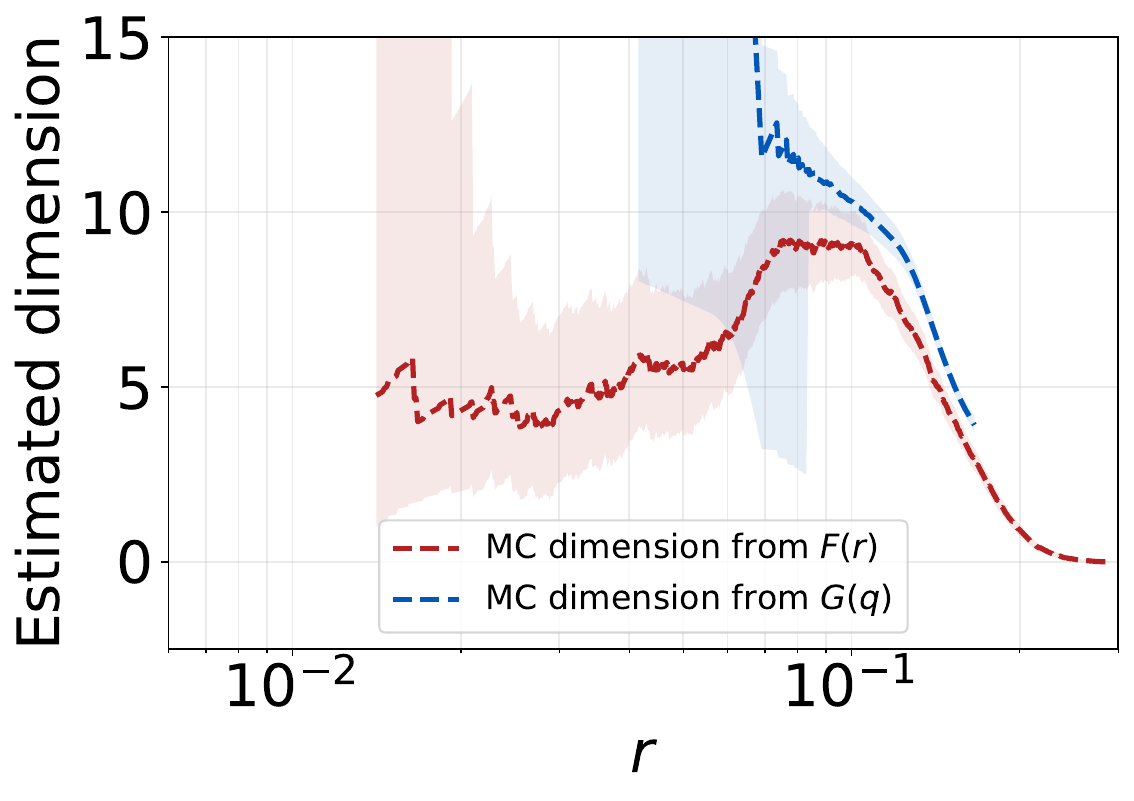}
    \end{subfigure}
    \hfill
    \begin{subfigure}{0.24\textwidth}
        \centering
        \includegraphics[width=\linewidth]{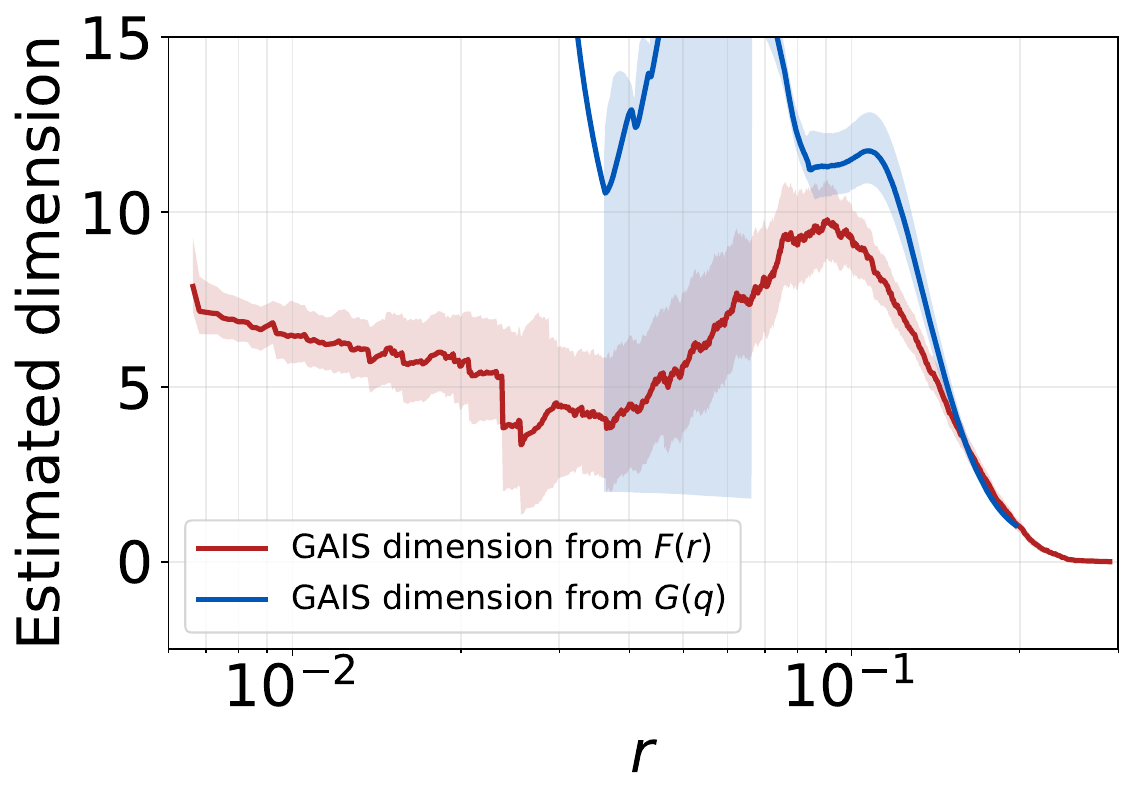}
    \end{subfigure}

    \caption{We compare the PDD dimension estimated using MC uniform sampling and our proposed GAIS. (first two plots: $5$-dimensional multivariate Gaussian; last two plots: MNIST).}
    \label{fig:PDD_dimension}
\end{figure*}

We show the PDD dimension in Figure~\ref{fig:PDD_dimension}. For Gaussian source, we study the $5$-dimensional multivariate Gaussian, where we know all the notions of dimension under study are equal to $d = 5$. We observe that, using the PDD estimated by MC uniform sampling (first plot), the CI becomes very large when $r$ is on the order of $10^{-1}$. By contrast, the $\underline{d}_Q(X)$ estimated using GAIS (second plot) has a narrow CI especially in the small $r$ region approaching $10^{-2}$. The the true dimension of the source is accurately estimated by the lower bound. Computation of the upper bound is less accurate, but shows similar trends. For MNIST, there is no ground truth dimension to compare with. However, comparing the third plot (Monte Carlo) and fourth plot (GAIS), we see that the use of importance sampling allows the dimension to be estimated with much tighter confidence intervals.

So, by utilizing the GAIS, we get a more stable and accurate dimension estimation for understanding the source. We note that, although GAIS provides a better $\overline{d}_Q(X)$ estimation compared with using MC uniform sampling, there is still room for improvement. One possible solution is to use inverse-variance weighting when computing $G(q) = \mathbb{E}_X[G(q;x)]$ empirically. We leave this as a interesting future direction for research.

\newpage

\end{document}